\documentclass{article}

 \usepackage[preprint]{neurips_2026}

\usepackage[utf8]{inputenc} % allow utf-8 input
\usepackage[T1]{fontenc}    % use 8-bit T1 fonts
\usepackage{hyperref}       % hyperlinks
\usepackage{url}            % simple URL typesetting
\usepackage{booktabs}       % professional-quality tables
\usepackage{amsfonts}       % blackboard math symbols
\usepackage{nicefrac}       % compact symbols for 1/2, etc.
\usepackage{microtype}      % microtypography
\usepackage{xcolor}         % colors

\usepackage{amsmath}
\usepackage{amssymb}
\usepackage{mathtools}
\usepackage{amsthm}
\usepackage{multirow}
\usepackage{mdframed}

\usepackage{hyperref}
\usepackage{url}
\usepackage{float}

\usepackage{array}
\usepackage{microtype}
\usepackage{graphicx}
\usepackage{xcolor}
\usepackage{amsmath}
\usepackage{afterpage}
\usepackage{bm}
\usepackage{dsfont}
\usepackage{mathtools}
\usepackage{nccmath}
\usepackage{amssymb}
\usepackage{multirow}
\usepackage{booktabs} % for professional tables
\usepackage{algorithm}
\usepackage{algpseudocode}
\usepackage{varwidth}
\usepackage{pifont}
\usepackage{makecell}
\usepackage[font=small,labelfont=bf,tableposition=top]{caption}
\usepackage{wrapfig}
\usepackage[flushleft]{threeparttable}
\usepackage{subcaption}
\usepackage{varwidth}
\usepackage{pifont}
\usepackage{makecell}
\usepackage{enumitem}
\usepackage{soul}
\usepackage{framed}
\usepackage[normalem]{ulem} % For underlining with \ul
\usepackage{listings}
\definecolor{keywordcolor}{rgb}{0.7, 0.1, 0.1}   % red
\definecolor{tacticcolor}{rgb}{0.0, 0.1, 0.6}    % blue
\definecolor{commentcolor}{rgb}{0.4, 0.4, 0.4}   % grey
\definecolor{symbolcolor}{rgb}{0.0, 0.1, 0.6}    % blue
\definecolor{sortcolor}{rgb}{0.1, 0.5, 0.1}      % green
\definecolor{attributecolor}{rgb}{0.7, 0.1, 0.1} % red
\lstdefinelanguage{Lean}{
  sensitive=true,
  morekeywords={
    theorem,lemma,example,def,by,have,show,from,where,let,in,if,then,else,
    match,with,fun,forall,exists,exact,import,open,namespace,section,end,
    variable,variables,inductive,structure,class,instance,noncomputable,
    norm_num,nlinarith,interval_cases,simp,simp_all,exact_mod_cast
  },
  morecomment=[l]{--},
  morecomment=[s]{/-}{-/},
  morestring=[b]",
  basicstyle=\ttfamily\small,
  keywordstyle=\color{keywordcolor},
  commentstyle=\color{commentcolor}\itshape,
  stringstyle=\color{black},
  showstringspaces=false,
  keepspaces=true,
  columns=fullflexible,
  breaklines=true
}
\setul{0.3ex}{0.6pt} % First argument: distance from text; second: underline thickness
\usepackage[capitalize,noabbrev]{cleveref}
\usepackage{tcolorbox}
\tcbuselibrary{listings,breakable}

\newtcblisting{promptbox}{
  listing engine=listings,
  listing only, breakable,
  colback=gray!5, colframe=black!75, arc=3mm, boxrule=0.5pt,
  left=2mm, right=2mm, top=1mm, bottom=1mm,
  listing options={
    language=,                              
    basicstyle=\ttfamily\small\color{black},
    numbers=none, frame=none,              
    keywordstyle=\color{black},
    commentstyle=\color{black},
    stringstyle=\color{black},
    identifierstyle=\color{black},
    showstringspaces=false,
    breaklines=true,
    breakatwhitespace=false,
    breakindent=0pt,                       
    columns=fullflexible,
    keepspaces=true,
    xleftmargin=0pt, xrightmargin=0pt,      
  }
}
\DeclareUnicodeCharacter{207B}{\textsuperscript{-}}

\theoremstyle{plain}

\theoremstyle{definition}

\theoremstyle{remark}

\usepackage[textsize=tiny]{todonotes}

\lstdefinestyle{mypython}{
    language=Python,
    basicstyle=\footnotesize\ttfamily,
    keywordstyle=\color{blue},
    commentstyle=\color{green!50!black},
    stringstyle=\color{red!70!black},
    numbers=left,
    numberstyle=\tiny,
    breaklines=true,
    showstringspaces=false,
    frame=single,
    extendedchars=true,
    inputencoding=utf8,
    literate=%
        {←}{{$\leftarrow$}}1
        {⁻¹}{{$^{-1}$}}2
}
\title{Lean Refactor: Multi-Objective Controllable\\Proof Optimization via Agentic Strategy Search}

\author{%
  Jialin Lu$^{1}$, Soonho Kong$^{2}$, Rodrigo Stehling$^{1}$, Kaiyu Yang$^{3}$,
  Zhangyang Wang$^{4}$, \\ \textbf{Weiran Sun}$^{1}$, \textbf{Wuyang Chen}$^{1}$ \\
  $^{1}$Simon Fraser University
  $^{2}$Amazon Web Services
  $^{3}$MiroMind \\
  $^{4}$University of Texas at Austin
}

\begin{document}

\maketitle

\begin{abstract}
We present \textbf{Lean Refactor}, a plug-and-play retrieval-augmented agentic framework for \emph{multi-objective, controllable, and version-robust} refactoring of Lean proofs. LLM-generated proofs are notoriously correct-but-verbose and brittle across library versions, yet existing refactoring works overlook \textbf{three practical challenges}: \textbf{1)} Lean refactoring is natively multi-objective (proof length, compilation cost, and version compatibility are often in tension); \textbf{2)} Lean repositories have fragile compatibility, whereas LLM releases are unaware of Lean/Mathlib versions; \textbf{3)} Training-based pipelines require repeated fine-tuning with each new LLM release, scaling neither with model churn nor with Lean's release cycle. Lean Refactor steers a \textbf{frozen} agentic LLM with retrievals from a curated database of \textbf{multi-objective refactoring strategies}, each densely annotated with metadata such as supported Lean/Mathlib versions and expected compilation-cost reduction. Experiments show over \textbf{70\% token-level compression} on competition benchmarks, over \textbf{20\%} on research repositories, and \textbf{up to 60\% compilation-time reduction}, outperforming prior work and Claude Code. Version-filtered retrieval further improves compression on the target Lean version, and refactored miniF2F proofs exhibit stronger \textbf{zero-shot version transfer} to future Lean releases than their unrefactored counterparts.
\end{abstract}
%for neurips: We will release our code, model, and data upon acceptance. !!!

\section{Introduction}

Lean and Mathlib~\cite{de2015lean,moura2021lean} enable collaborative, machine-verified mathematics and provide a strong reinforcement learning testbed that grounds large language model (LLM) reasoning in formal proofs, supporting gold-medal International Mathematical Olympiad (IMO) performance~\cite{trinh2024solving,achim2025aristotle} and rapid progress in LLM-powered provers~\cite{alphaproof,xin2024deepseek,xin2024deepseekv15,ren2025deepseek,lin2025goedel,lin2025goedelproverv2scalingformaltheorem}. However, LLMs trained with reinforcement learning (RL) often produce Lean proofs that are correct but \textbf{notoriously long and hard to read}~\cite{ahuja2024improver,gu2025proofoptimizer}. Because RL rewards target proof correctness, models are insensitive to redundant steps or heavy automation where simple steps suffice. Such proofs are difficult for humans to understand and computationally expensive to compile, posing particular problems for large libraries like Mathlib.

Recent work on LLM-based code refactoring~\cite{gautam2025refactorbench,gong2025language} pursues either fine-tuning LLMs~\cite{gu2025proofoptimizer,yang2025perfcoder,gong2025tuning} or agentic frameworks built on pretrained general-reasoning LLMs~\cite{ahuja2024improver,depalma2024exploring,ospanov2025apollo,piao2025refactoring,oueslati2025refagent,karabiyik2025refactorgpt,he2025swe,cui2025large,bai2025polo,zhao2025semopt,wu2025fasterpy}. However, LLMs have seen limited adoption for refactoring Lean proofs,
and we point out \textbf{three key limitations}.
\textbf{\emph{First, Lean refactoring is natively multi-objective.}}
Lean code is often simultaneously optimized along several axes, such as proof length and compilation cost. These objectives are entangled and frequently in tension: shorter proofs may invoke heavier tactics that inflate compilation cost, while cheaper proofs may sprawl in length. Standard pretraining, RLHF, and fine-tuning~\cite{ouyang2022training,achiam2023gpt} offer no principled mechanism to balance such trade-offs, let alone the \textbf{unseen objectives} users routinely introduce at inference time.

\begin{wrapfigure}{r}{0.5\textwidth}
\vspace{-3em}
\centering
\includegraphics[width=1.0\linewidth]{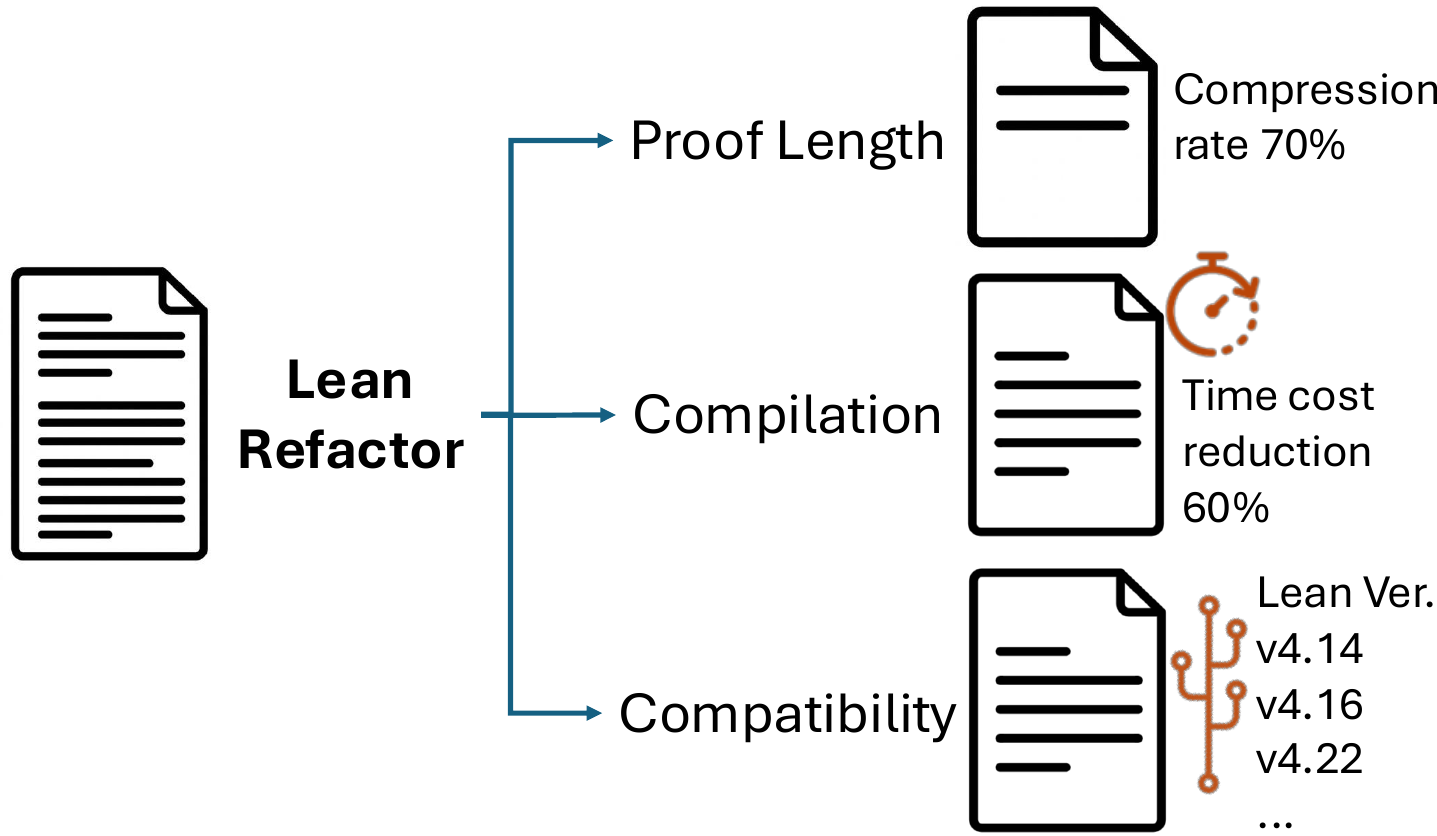}
\centering 
\vspace{-1.5em}
\captionsetup{font=small}
\caption{Our Lean Refactor can be controlled to refactor Lean proofs for multiple objectives: proof length, compilation time cost, and compatibility across Lean versions.}
\vspace{-2em}
\label{fig:teaser}
\end{wrapfigure}

\textbf{\emph{Second, frontier LLMs are misaligned with the Lean/Mathlib release cycle.}} Lean and Mathlib evolve on a near-weekly basis, with lemmas renamed, APIs restructured, and new automation landed, while LLMs are pinned to a knowledge cutoff. Even with web-search augmentation, \textbf{no principled way exists for maintainers to pair a codebase version with a compatible model}, leading to pervasive hallucination of stale tactics and removed APIs.

\textbf{\emph{Third, paired refactoring data is scarce.}} Compared with mainstream languages (C/C++, Java, Python), Lean has a far smaller codebase, and existing Mathlib and community-aggregated competition corpora~\cite{minif2f,putnam,li2024numinamath} provide neither refactoring trajectories nor noisy-clean code pairs.

To address these fundamental bottlenecks, we investigate our core question:
\vspace{-0.7em}
\begin{center}
\fbox{
    \parbox{0.9\linewidth}{
        \textit{Can a frozen LLM be steered at inference time 
    to balance competing refactoring objectives while reliably refactoring code in fast-evolving Lean/Mathlib environments?}
    %     \textit{\textbf{Q2:} Can a frozen LLM reliably refactor code in 
    % hyper-evolving Lean/Mathlib environments?} \\
        % \textit{\textbf{Q2:} Without large-scale noisy-to-clean proof pairs, 
        % can we distill reusable refactoring strategies that transfer across 
        % proof domains?}
    }
}
\end{center}
\vspace{-0.7em}

In this paper, we introduce \textbf{Lean Refactor}, a retrieval-augmented agentic framework that externalizes refactoring knowledge into a structured database. Lean Refactor grounds an agentic LLM with a curated corpus of \textbf{multi-objective, maintainable refactoring strategies}. Each retrieved strategy supplies the LLM with aligned structural context and explicit execution instructions (\emph{when to apply}, \emph{how to apply}, and \emph{examples}). Crucially, these strategies are \textbf{densely annotated with refactoring metadata}, tracking signals such as expected compilation-cost reduction and validated Lean/Mathlib version compatibility. By dynamically filtering and reranking these annotated strategies at inference time, our framework steers proof optimization toward user-specified objectives without model fine-tuning, and remains version-robust as Lean/Mathlib evolves beyond the LLM's knowledge cutoff.

We summarize our contributions below:
\begin{enumerate}[leftmargin=*]
    \item We release the largest Lean refactoring dataset to date: over 200K long-short proof pairs and 9K refactoring strategies, each annotated with compilation-cost reduction and compatible Lean/Mathlib versions.
    \item Lean Refactor achieves over 70\% compression on competition proofs (miniF2F~\cite{minif2f}, PutnamBench~\cite{putnam}, Putnam2025) and over 20\% on research repositories (PFR~\cite{pfr_formalization}, PhysLean~\cite{physlib_lean}, Analysis~\cite{tao_lean_analysis}, FLT~\cite{FLT_Lean}), outperforming prior work and Claude Code while remaining compatible with Gemini, Claude, and GPT series backbones.
    \item Lean Refactor supports multi-objective, controllable refactoring: beyond length, it cuts compilation cost by over 30\%, holds compression steady across Lean versions, and produces miniF2F proofs that type-check across more future Lean/Mathlib releases than their unrefactored counterparts.
\end{enumerate}

% \section{Challenges of LLM-based Maintainable Lean Refactoring}
\section{Lean Refactoring is Challenging for LLMs}
\label{sec:refactoring_strategies}

\begin{wrapfigure}{r}{0.4\textwidth} % {r}
\vspace{-5em}
\centering
\includegraphics[width=1\linewidth]{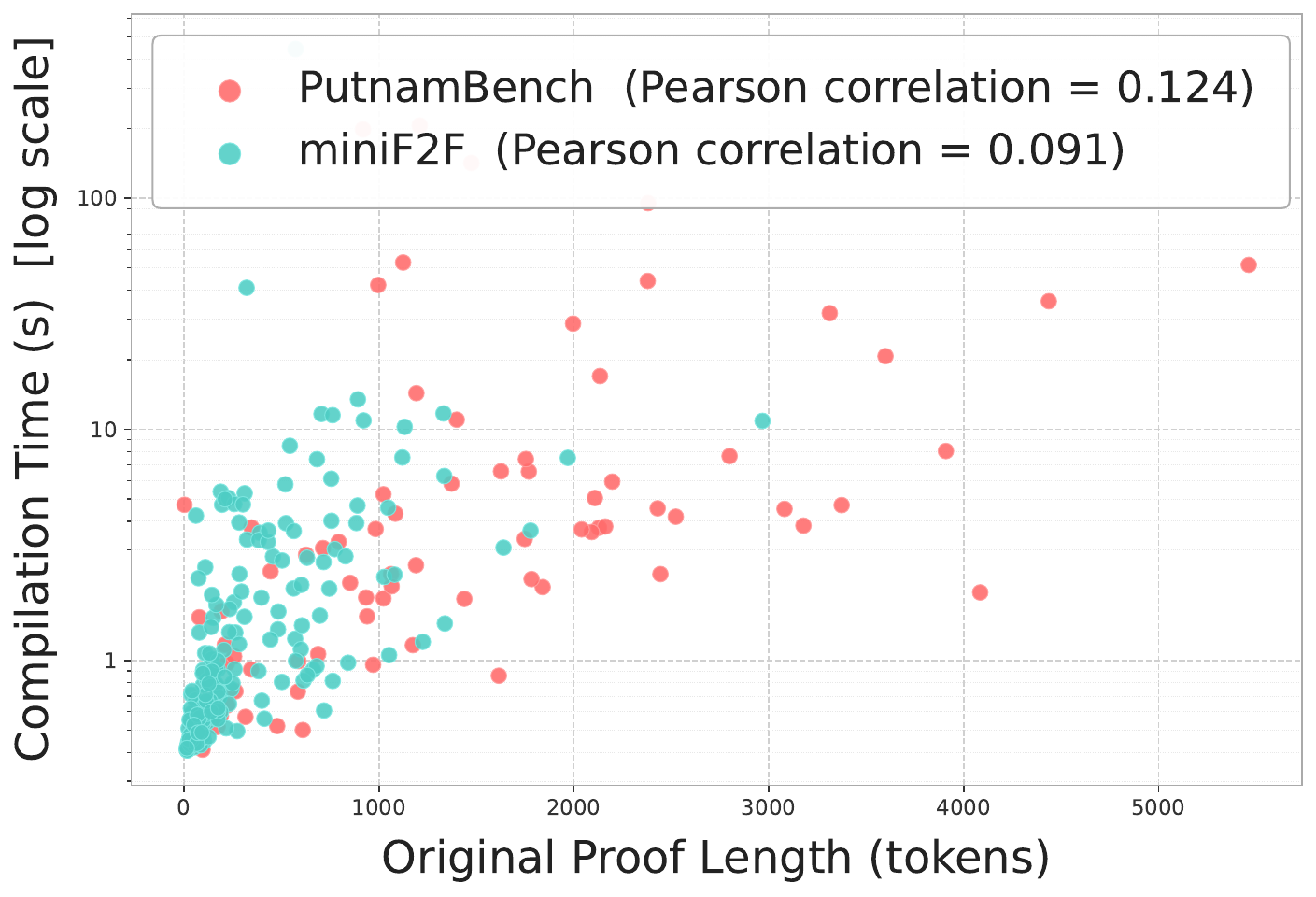}
\vspace{-1.5em}
\captionsetup{font=small}
\caption{\textbf{Weak correlation} between proof length and compilation time. Token length fails to predict compilation cost because of short, heavy tactics, a factor often overlooked in previous Lean refactoring research.
% Top: give an intuitive example and reason why proof length and compilation time are not fully aligned.
}
\label{fig:proof_length_time_demo}
\vspace{-2em}
\end{wrapfigure}

Beyond the redundancy in code length demonstrated in previous papers~\cite{ahuja2024improver,gu2025proofoptimizer},
% which we will not re-iterate here.
in this section, we motivate our work with more practical challenges of Lean code refactoring with LLMs.

\paragraph{Proof Optimization is Multi-Objective.} While proof length is the most convenient refactoring objective, as it can be computed during LLM inference, compilation cost is an equally important complexity metric, and the two are not always aligned (Figure~\ref{fig:proof_length_time_demo}; see Appendix~\ref{sec:len_vs_time} for an example). Prior Lean refactoring work either ignores compilation cost or swaps it in as a substitute inference-time objective, rather than jointly optimizing it with length~\cite{ahuja2024improver,gu2025proofoptimizer}. More broadly, while recent formal-math post-training increasingly relies on verifiable rewards (e.g., compiler-judged correctness), continuous runtime signals such as \textbf{compilation cost remain underused}~\cite{lin2025goedelproverv2scalingformaltheorem, gu2025proofoptimizer}.

\begin{wrapfigure}{r}{0.5\textwidth} % {r}
\vspace{-2em}
\centering
\includegraphics[width=1.0\linewidth]{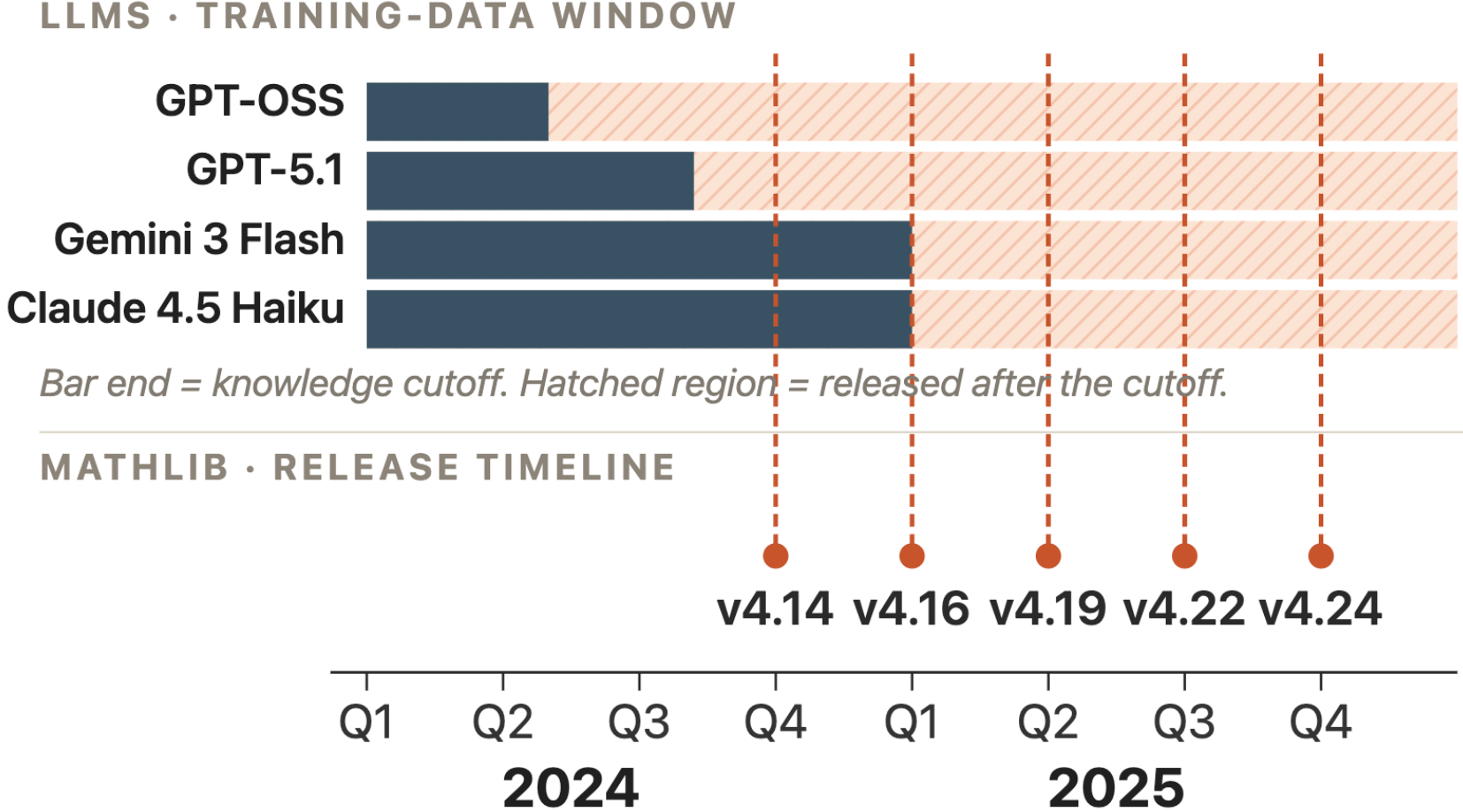}
\vspace{-1.5em}
\centering 
\captionsetup{font=small}
% https://chatgpt.com/c/6985b8fb-dda8-832f-8edf-825b14b3c062
\caption{Knowledge cutoff lag in LLMs: Most LLMs misalign with Lean/Mathlib versions due to static knowledge cutoff.
% In reinforcement learning, rewards for LLMs are also unaware of compilation time costs of different Lean code~\cite{ouyang2022training,achiam2023gpt}.
% Public reports of LLM indicate post-training via reinforcement learning mainly leverages rewards from learned preference or rubric models over text outputs, rather than external runtime metrics like Lean compiler compilation-time cost~\cite{ouyang2022training,achiam2023gpt}.
}
\vspace{-1em}
\label{fig:repo_compatibility}
\end{wrapfigure}

\paragraph{Lean Repository Compatibility is Fragile.} Version sensitivity is a long-standing bottleneck of Lean/Mathlib: Lean and Mathlib evolve rapidly; weekly releases frequently rename declarations and break imports even across patch versions~\cite{baanen2025growing}, forcing downstream projects to pin to Mathlib's exact toolchain~\cite{mathlib_breakage}\footnote{In a Zulip thread on mathlib-dependent projects, the requirement is stated bluntly: ``It needs to be the exact same version.'' See \href{https://leanprover-community.github.io/archive/stream/270676-lean4/topic/mathlib4.html}{this discussion}.}. Agentic LLMs, however, are \textbf{never aligned} with a specific version: as Figure~\ref{fig:repo_compatibility} shows, every recent frontier model's training cutoff lands \emph{before} recent Mathlib releases, leaving newer APIs unseen and the model's coverage of older versions opaque. This pushes maintainers into continuous, costly upgrades. \textbf{Lean Refactor} mitigates this misalignment by grounding the agent in a strategy bank whose strategies carry explicit Mathlib-version metadata: retrieval is filtered to the user's target toolchain, and the bank itself can be re-profiled as new Mathlib versions ship.

\paragraph{Competitions Fail to Reflect Research-Level Proofs.} Current LLM provers are primarily trained on competition mathematics, such as NuminaMath~\cite{li2024numinamath} and Goedel-Pset~\cite{lin2025goedel}. However, competition proofs do not reflect the structural complexity of research-level Lean code. As Table~\ref{table:theorem_used} shows, research proofs rely far more heavily on Mathlib and intra-project dependencies. This dense network of external and internal lemmas makes research-level code harder to refactor, forcing the model to reason over a much larger and less familiar context to find valid simplifications. Consequently, despite their practical importance, research-level proofs across diverse domains have been largely overlooked in prior Lean proof optimization work~\cite{gu2025proofoptimizer, ahuja2024improver}.

\section{Methods}

\begin{figure*}[t]
% \vspace{-1.2em}
\centering
\includegraphics[width=1.\linewidth]{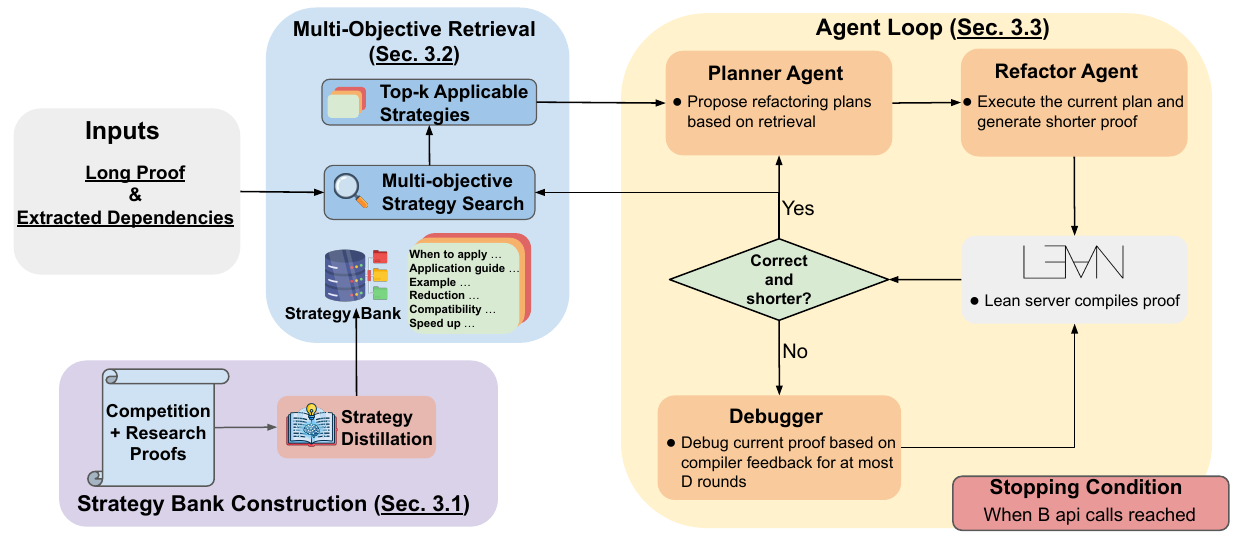}
\centering 
\vspace{-1.5em}
\captionsetup{font=small}
\caption{Overview of Lean Refactor framework.
1. We first summarize raw long-short proof pairs sourced from diverse Lean code sources into well-structured reusable strategies, and more importantly, we annotate each strategy with code refactoring metadata, including compilation time cost and version compatibility (Section~\ref{sec:strategy_bank}).
2. We then train a strategy retrieval model that maps Lean code (that can be potentially optimized) to one or more refactoring strategies, with multi-objective re-rank and filtering (Section~\ref{sec:multi_objective_retrieval}).
3. Finally, we design an agent workflow that intelligently plans refactoring, generates shortened proofs, and debugs (Section~\ref{sec:agent_description}).
Components in our agent workflow (planner, refactor, debugger) are frontier LLMs.}
\vspace{-2em}
\label{fig:framework_overview}
\end{figure*}

Lean Refactor is a retrieval-augmented agentic framework built around three design choices that directly address the bottlenecks in Section~\ref{sec:refactoring_strategies}. We externalize refactoring knowledge into a densely annotated strategy bank whose metadata, including Lean/Mathlib version compatibility and expected compilation cost reduction, mitigates the misalignment between the LLM's training cutoff and Lean/Mathlib release cycles (Section~\ref{sec:strategy_bank}). At inference, the target proof is semantically grounded via dependency annotations~\footnote{We use adapted versions of jixia and ntp-toolkit for dependencies extraction.}, while a multi-objective retrieve-and-rerank pipeline steers the agent toward user-specified trade-offs between proof length, compilation cost, and version compatibility (Section~\ref{sec:multi_objective_retrieval}). We wrap these components in a plug-and-play agent loop whose planner, refactorer, and debugger are instantiated by any frozen frontier LLM, so the framework transfers across Gemini, Claude, and GPT series without any retraining (Section~\ref{sec:agent_description}). Figure~\ref{fig:framework_overview} illustrates the pipeline; pseudocode is in Algorithm~\ref{alg:agent_framework}.

\subsection{A Densely Annotated, Version-Aware Strategy Bank}
\label{sec:strategy_bank}

\paragraph{Strategy structure.} Each entry is a reusable refactoring pattern distilled from a raw long--short proof pair, organized into six fields: \emph{title}, a natural-language \emph{description}, a precondition pattern (\emph{when to apply}), a step-by-step \emph{application guide}, a \emph{before/after example}, and a categorical \emph{potential length reduction} tag. The schema preserves enough structural context for the LLM to generalize, promoting transfer to unseen proofs. Examples appear in Appendix~\ref{sec:strategy_examples}.

\paragraph{Refactoring metadata.} Each strategy is additionally annotated along two axes that the LLM cannot infer from its training data:
\begin{itemize}[leftmargin=*]
    \item \emph{Compilation-time reduction.} We profile both proofs in every long--short pair with \texttt{lake env lean --profile}, isolating proof execution from import overhead, and record the relative speedup. Since each strategy can be instantiated by many long--short pairs, we keep the median speedup across all pairs associated with the strategy as the strategy-level estimate.
\item \emph{Version compatibility.} For all non-Mathlib pairs (Mathlib is locked to its toolchain), we recompile each shortened proof under three additional Lean toolchains beyond the original v4.24.0: v4.14.0, v4.16.0, and v4.22.0. A strategy's compatibility set is the intersection of toolchains under which all of its shortened proofs continue to compile, certifying that the transformation is robust across the interval rather than merely syntactically plausible in a single snapshot. This metadata is what makes the bank version-aware at inference time (Section~\ref{sec:multi_objective_retrieval}) and is the primary lever by which we mitigate the misalignment between the LLM's training cutoff and Lean/Mathlib release cycles.\end{itemize}

\paragraph{Bank construction.} We aggregate statements from NuminaMath-1.5~\cite{li2024numinamath}, FineLeanCorpus~\cite{peng2025criticleancriticguidedreinforcementlearning}, Mathlib, and ATLAS~\cite{liu2025atlasautoformalizingtheoremslifting}, and synthesize proofs with Goedel-Prover-V2-32B~\cite{lin2025goedelproverv2scalingformaltheorem} and GPT-OSS-120B~\cite{openai2025gptoss120bgptoss20bmodel}. We decontaminate the aggregated theorems against our held-out evaluation set (defined in Section~\ref{sec:experiment}) via exact-match and embedding-based filtering to preclude leakage. Pairs are built by shortening long proofs and synthetically expanding short ones. GPT-OSS-120B distills each pair into one or more location-grounded strategies; a second LLM-as-judge pipeline filters incorrect extractions; and a Qwen3-Embedding-8B~\cite{zhang2025qwen3embeddingadvancingtext} based iterative de-duplication pipeline clusters semantically equivalent strategies. The final corpus contains \textbf{9237 unique strategies} distilled from 481567 raw strategy extractions over 200K long--short proof pairs. Full details are in Appendix~\ref{sec:strategy_bank_details}.

\subsection{Multi-Objective Retrieval for Controllable Refactoring}
\label{sec:multi_objective_retrieval}
\paragraph{Retrieval model training.} We fine-tune Qwen3-Embedding-8B~\cite{zhang2025qwen3embeddingadvancingtext} on 639090 (query, strategy) pairs. Our training procedure largely follows prior Lean code search work~\cite{lu2025lean}, with several adaptations tailored to our refactoring setting. During strategy distillation from long--short proof pairs, we record the code segment in the long proof on which the strategy operates (Appendix~\ref{app:sb-summarization}); we use this segment as the query and the strategy's \emph{when-to-apply} clause as its positive target. Let $B$ denote the mini-batch size, $q_i$ the $i$-th query, $c_i^+$ its ground-truth positive strategy, and $\mathcal{C}_B = \{c_j^+\}_{j=1}^{B}$ the set of positives in the batch, where each $c_j^+$ with $j \neq i$ serves as an in-batch negative for query $q_i$. We write $\mathrm{Sim}(\cdot,\cdot)$ for cosine similarity, $\tau$ for the softmax temperature, and $m$ for a false-negative margin. The indicator $\mathbb{I}(q_i, c)$ equals $0$ when $\mathrm{Sim}(q_i, c) > \mathrm{Sim}(q_i, c_i^+) + m$, masking out candidates that score higher than the positive by more than $m$, and $1$ otherwise. The training objective is
\begin{equation}
    \mathcal{L} = -\frac{1}{B}\sum_{i=1}^{B} \log \frac{\exp(\mathrm{Sim}(q_i, c_i^+)/\tau)}{\sum_{c \in \mathcal{C}_B} \mathbb{I}(q_i, c)\,\exp(\mathrm{Sim}(q_i, c)/\tau)}.
\end{equation}
We additionally apply boundary augmentation, jittering segment boundaries to bridge the gap between clean training spans and noisy runtime inputs. Training details are in Appendix~\ref{sec:retrieval_model_training_details}.

\begin{figure}[t!]
% \vspace{-1.2em}
\centering
\includegraphics[width=1.\linewidth]{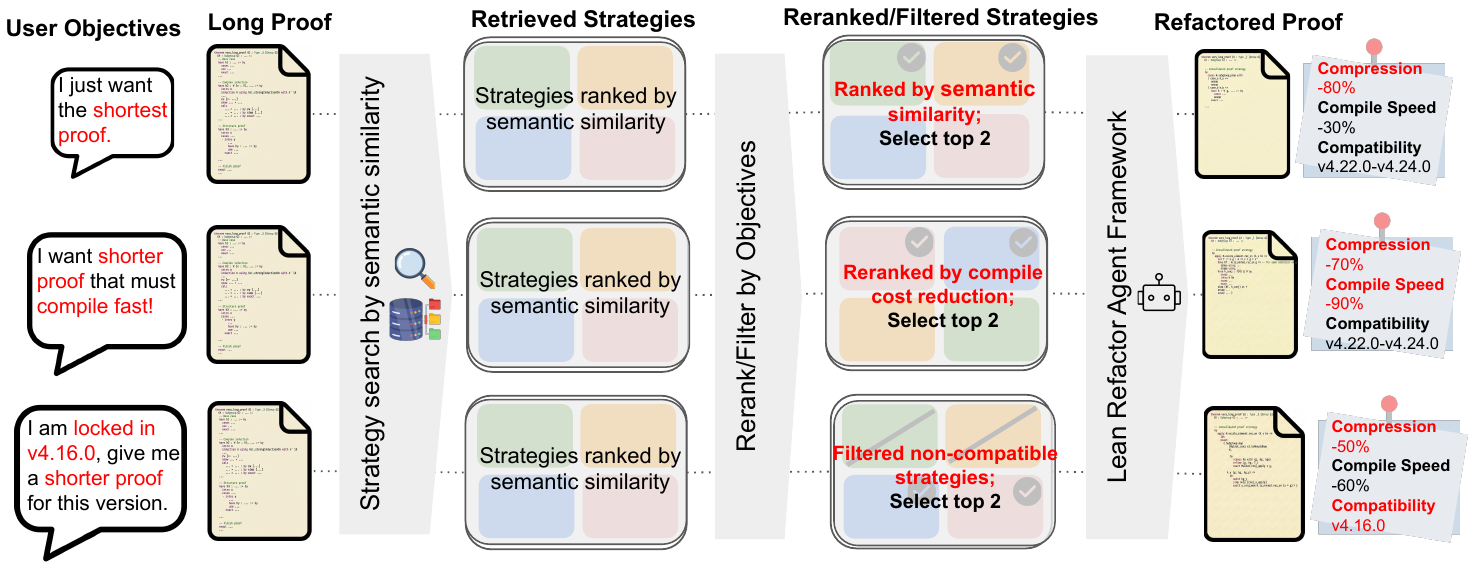}
\centering 
\vspace{-1.5em}
\captionsetup{font=small}
\caption{\textbf{Multi-objective Lean proof refactoring.} A single strategy bank adapts to diverse user objectives at inference without retraining. \emph{Top:} for shortest proofs, top-$K$ strategies by cosine similarity are used directly. \emph{Middle:} to balance faster compilation, candidates are reranked by annotated compile-cost reduction. \emph{Bottom:} for a target toolchain (e.g., v4.16.0), incompatible strategies are filtered out. The agent outputs (right) reflect these distinct trade-offs. Metadata-driven objectives are fully composable and easily extensible to new criteria.}
\vspace{-1.5em}
\label{fig:multi_objective}
\end{figure}

\paragraph{Objective-conditioned retrieval at inference (Figure~\ref{fig:multi_objective}).} The user specifies an objective at inference time, and we reshape the retrieval rule accordingly while operating over the \emph{same} underlying bank:
\begin{itemize}[leftmargin=*]
    \item \emph{Shortest proof.} If the user prioritizes proof length, we use pure cosine-similarity top-$K$, exposing the agent to the most structurally relevant strategies.
    \item \emph{Compilation-time aware.} If the user prioritizes compile-time reduction, we apply a two-stage retrieve-and-rerank procedure: we fetch a broad pool of semantically similar candidates, then rerank in descending order of annotated compile-cost reduction. Because the context window is bounded, top-$K$ by similarity alone might exclude high-efficiency strategies; reranking surfaces them without sacrificing applicability.
    \item \emph{Version-specific.} If the user targets a particular Lean version, we filter the candidate pool to strategies whose compatibility set includes that version, removing version-incompatible strategies.
\end{itemize}
Each objective is realized as a different retrieval rule over the same shared bank. This design composes cleanly: version filtering can be stacked on top of compile-time reranking to jointly target both objectives. It also admits \emph{new} objectives at inference time by introducing new metadata fields without touching the model.

\subsection{A Plug-and-Play, LLM-Agnostic Agent Loop}
\label{sec:agent_description}
\paragraph{Iterative agent loop (Algorithm~\ref{alg:agent_framework}, Figure~\ref{fig:framework_overview}; see Appendix~\ref{sec:prompts} for prompts).} Given a long proof, we annotate it with the signatures of all invoked dependencies to provide semantic grounding. The annotated proof then enters an agentic loop that coordinates four distinct stages:
\begin{enumerate}[leftmargin=*]
    \item \emph{Retrieval.} We segment the annotated proof at multiple granularities to enable refactorings of varying scope. Each segment is embedded and used to query the strategy bank; the top-$K$ strategies are retrieved and processed under the user-specified objective (Section~\ref{sec:multi_objective_retrieval}), then passed forward with their target locations.
    \item \emph{Planner.} A frozen LLM consumes the current proof, the retrieved strategies, and execution/planning history, and decomposes the task into a sequence of refactoring steps mapped to specific segments (or the entire proof).
    \item \emph{Refactorer.} The same LLM, given one planned step, rewrites the targeted segment (or the entire proof). The candidate is handed to the Lean 4 compiler for validation.
    \item \emph{Debugger.} On compile failure, compiler feedback is routed back to the LLM for up to $D$ rounds of localized repair, constrained to the planner's original intent.
\end{enumerate}
On successful compilation \emph{and} length reduction, the updated proof and trace are fed back to the planner for dynamic replanning; on persistent failure, the step is logged and skipped. The loop terminates at a budget of $B$ LLM calls or when proof length falls below target $T$. 

\paragraph{Adaptability and Sustainability.} Our design choices keep the framework agnostic to both LLM backbones and Lean versions. The planner, refactorer, and debugger are model-agnostic agent roles that operate over any frozen frontier LLM, so the framework can transfer to a new backbone with no fine-tuning or adaptation step. Refactoring knowledge, including strategies together with their version and compilation metadata, lives in an external bank that is retrieved at inference time and can be updated as Lean/Mathlib evolves. New optimization objectives are supported by acquiring the corresponding metadata, with no retraining required. As a result, new frontier LLMs inherit the bank's full coverage immediately, and new Lean versions are supported by re-profiling the bank. Section~\ref{sec:experiment} empirically validates interchangeable deployment across Gemini, Claude, and GPT-OSS.

\section{Experiments}
\label{sec:experiment}

\paragraph{Benchmarks.} We evaluate Lean Refactor on seven benchmarks across two regimes. \emph{Competition:} miniF2F~\cite{minif2f} (194 theorems), PutnamBench~\cite{putnam} (75 theorems), and Putnam2025 (66 theorems from problems B1/B2 of Seed-Prover~1.5~\cite{chen2025seedprover15masteringundergraduatelevel}). For miniF2F and PutnamBench, we reuse the long proofs released by ProofOptimizer~\cite{gu2025proofoptimizer} so both methods refactor identical inputs. \emph{Research:} 45 theorems sampled from each of Analysis~\cite{tao_lean_analysis}, FLT~\cite{FLT_Lean}, PFR~\cite{pfr_formalization}, PhysLean~\cite{physlib_lean}, and PDE~\cite{Stehling2026Lean}, along with our 15 generated solutions to Verina~\cite{ye2026verinabenchmarkingverifiablecode}, spanning diverse proof lengths with all internal and external dependencies retained.

\paragraph{Setup and metrics.} Unless noted otherwise, all runs use Gemini~3~Flash~\cite{geminiteam2025gemini3flash}, Lean v4.24.0, an API call budget of 30, and the syntax-aware tokenizer from ProofOptimizer (code in Appendix~\ref{sec:proof_length_code}). We report three primary metrics: \textbf{proof length reduction} (Section~\ref{sec:results_proof_length}), \textbf{compilation-time reduction} (Section~\ref{sec:results_compilation_time}), and \textbf{cross-version compatibility} (Section~\ref{sec:results_compatibility}). Heartbeat reduction and additional compilation-cost results are deferred to Appendices~\ref{sec:heartbeat_comparison} and~\ref{sec:compilation_time_research_level}; dataset statistics, length distributions, and full experimental details to Appendix~\ref{sec:experimental_settings}.

\paragraph{Baselines.} On miniF2F and PutnamBench, we compare against \textbf{ProofOptimizer}~\cite{gu2025proofoptimizer} using their released optimized proofs\footnote{ProofOptimizer's released proofs are produced via an 8-iteration shortening process with sampling budgets of $6{\times}64+2{\times}1024$ per theorem (64 samples per iteration for the first 6, 1024 for the last 2) from a fine-tuned 7B model, while our system uses 30 API calls per theorem orchestrating planner, refactorer, and debugger steps over a frozen frontier LLM. The two paradigms differ in model scale and sampling-vs-agentic control flow, and since ProofOptimizer's checkpoints are not publicly available, a matched-backbone or matched-budget comparison is not possible; we therefore evaluate each system under its released configuration. Our ablations (Tables~\ref{tab:proof_length_competition_results} and ~\ref{tab:proof_length_research_results}) isolate framework contributions on a shared backbone, and we additionally compare against Claude Code under matched budget and backbone (Table~\ref{tab:claude_haiku_combined}).}; their closed-source checkpoints preclude evaluation on the remaining benchmarks. We additionally isolate each component of our framework via four internal ablations: \textbf{Base Agent} (refactorer + debugger + retrieval, planner removed), \textbf{Base Agent w/o Retrieval} (planner and retrieval removed), \textbf{Lean Refactor w/o Retrieval} (planner + refactorer + debugger, retrieval removed), and \textbf{Lean Refactor w/ Random Retrieval} (retrieval replaced by sampling over the bank).

To test LLM-agnosticism, we re-run our framework with Claude Haiku~4.5~\cite{anthropic2025claude45haiku} and compare against \textbf{Claude Code} (also powered by Haiku~4.5), giving Claude Code a dollar budget matched to our framework's Haiku 4.5 runs. We evaluate Claude Code under three tool configurations of increasing tool access: (i) Lean MCP\footnote{\url{https://github.com/oOo0oOo/lean-lsp-mcp}} only, providing compiler feedback; (ii) Lean MCP + Lean skills\footnote{\url{https://github.com/cameronfreer/lean4-skills}}, adding general Lean~4 refactoring and golfing guidance; and (iii) the above plus our strategy retrieval tool, which exposes the strategy bank as an on-demand resource the agent can query during refactoring. Due to budget, this comparison is restricted to one representative benchmark per regime: PutnamBench (competition) and Analysis (research-level). See Appendix~\ref{sec:experimental_settings} for further details.

\subsection{Proof Length Reduction}
\label{sec:results_proof_length}

% \vspace{-1.5em}
\begin{table}[ht]
\centering
\caption{\textbf{Average relative proof-length reduction (\%) on competition benchmarks using Gemini 3 Flash.} ``--'' indicates the baseline could not be evaluated. Best results in bold.}
\label{tab:proof_length_competition_results}
\resizebox{!}{1.5cm}{%
\begin{tabular}{lccc}
\toprule
\textbf{Method}  & \textbf{miniF2F~\cite{minif2f}} & \textbf{PutnamBench~\cite{putnam}} & \textbf{Putnam 2025} \\
\midrule
ProofOptimizer~\cite{gu2025proofoptimizer} &  87.90 & 57.20 & -- \\
\midrule
Base Agent                            & 79.75 & 41.15 & 68.18 \\
Base Agent w/o Retrieval             & 79.37 & 37.02 & 63.91 \\
Lean Refactor w/o Retrieval          & 83.44 & 62.03 & 75.22 \\
Lean Refactor w/ Random Retrieval    & 83.72 & 59.41 & 74.29 \\
\textbf{Lean Refactor}               & \textbf{88.40} & \textbf{70.38} & \textbf{79.27} \\
\bottomrule
\end{tabular}
}
\vspace{-1em}
\end{table}

\begin{table}[ht]
\centering
\caption{\textbf{Average relative proof-length reduction (\%) on research-level benchmarks using Gemini 3 Flash.} Best results in bold.}
\label{tab:proof_length_research_results}
\resizebox{!}{1.3cm}{%%
\begin{tabular}{lccccc}
\toprule
\textbf{Method} & \textbf{Analysis~\cite{tao_lean_analysis}} & \textbf{FLT~\cite{FLT_Lean}} & \textbf{PFR~\cite{pfr_formalization}} & \textbf{PhysLean~\cite{physlib_lean}} \\
\midrule
Base Agent                        & 17.22 & 12.36 & 8.72  & 10.82 \\
Base Agent w/o Retrieval          & 12.10 & 11.39 & 7.93  & 10.90 \\
Lean Refactor w/o Retrieval       & 26.54 & 19.72 & 18.83 & 18.74 \\
Lean Refactor w/ Random Retrieval & 24.46 & 21.47 & 18.91 & 23.21 \\
\textbf{Lean Refactor}            & \textbf{33.54} & \textbf{23.55} & \textbf{20.37} & \textbf{27.69} \\
\bottomrule
\end{tabular}
}
\vspace{-2.em}
\end{table}

Tables~\ref{tab:proof_length_competition_results} and~\ref{tab:proof_length_research_results} report length reductions with Gemini~3~Flash (test-time scaling curves in Appendix~\ref{sec:tts_figures}); Table~\ref{tab:claude_haiku_combined} reports the LLM-agnostic check with Haiku~4.5 and comparison with Claude Code. Additional results with PDE~\cite{Stehling2026Lean} and Verina~\cite{ye2026verinabenchmarkingverifiablecode} are shown in Appendix~\ref{sec:extra_length_reduction_results}. \textbf{Lean Refactor achieves state-of-the-art performance across both regimes:} on competition benchmarks it surpasses ProofOptimizer, with the gap widening on PutnamBench, where longer proofs benefit most from retrieval-augmented planning, and on human-written research-level proofs it achieves substantial additional compression and beats every ablation. \textbf{Planning and retrieval are synergistic.} The planner is critical because without it, the Base Agent becomes overwhelmed by long proofs and stalls in persistent debug loops. Conversely, retrieval only pays off when the planner explicitly sequences the provided strategies. Results show that random retrieval yields inconsistent improvements, whereas targeted retrieval delivers consistent and substantially larger gains. This confirms that the framework's success stems from \emph{accurate strategy selection} rather than mere context augmentation.

\vspace{-1.em}
\begin{table}[ht]
\centering
\caption{\textbf{Average relative proof-length reduction (\%) on PutnamBench and Analysis with Claude Haiku 4.5, compared with Claude Code.} We evaluate our framework and ablation variants with Claude Haiku 4.5 as the backbone and compare against Claude Code (also powered by Claude Haiku 4.5) under three tool configurations: Lean MCP only, Lean MCP with Lean skills, and Lean MCP with Lean skills plus our strategy retrieval tool. Budget is matched by dollar cost. Best result in bold.}
\label{tab:claude_haiku_combined}
\resizebox{\linewidth}{!}{%
\begin{tabular}{llcc}
\toprule
\textbf{Agent} & \textbf{Configuration} & \textbf{PutnamBench~\cite{putnam}} & \textbf{Analysis~\cite{tao_lean_analysis}} \\
\midrule
\multirow{3}{*}{\shortstack[l]{Claude Code\\(Claude Haiku 4.5)}}
 & Lean MCP only                                  & 33.71 & 7.86  \\
 & \quad + Lean skills                   & 33.29 & 8.83  \\
 & \quad\quad + Our strategy retrieval tool                & 34.91 & 10.05 \\
\midrule
\multirow{5}{*}{\shortstack[l]{Ours\\(Claude Haiku 4.5)}}
 & Base Agent                                     & 39.21 & 15.45 \\
 & Base Agent w/o Retrieval                       & 36.89 & 10.88 \\
 & Lean Refactor w/o Retrieval                    & 48.76 & 18.33 \\
 & Lean Refactor w/ Random Retrieval              & 46.94 & 18.59 \\
 & \textbf{Lean Refactor}                         & \textbf{51.35} & \textbf{21.09} \\
\bottomrule
\end{tabular}
}
\vspace{-1em}
\end{table}

\textbf{The framework is LLM-agnostic and the strategy bank improves the performance of general-purpose agents.} The same ranking holds when the backbone is switched to Claude Haiku~4.5 (Table~\ref{tab:claude_haiku_combined}), and under a matched budget, Claude Code itself improves on both benchmarks once given access to our strategy retrieval tool. Even so, our full framework outperforms the strongest Claude Code configuration, demonstrating that our tailored planner and compiler-in-the-loop architecture provides targeted guidance that a general-purpose agent lacks, and more effectively leverages strategy retrieval.  Beyond raw length, refactored proofs also reduce external and intra-project dependencies in 42--67\% of cases across all benchmarks, evidence that the compression reflects genuine logical simplification rather than offloading work onto heavy library lemmas, which further improves readability and version resilience (Appendix~\ref{sec:results_dependencies}).

\subsection{Multi-Objective Control: Compilation Time}
\label{sec:results_compilation_time}

\begin{table}[ht]
\centering
\caption{\textbf{Compilation-time reduction on miniF2F, PutnamBench, and Putnam2025.} \emph{Avg. Rel. Reduc.}: average relative \% decrease in compile time. \emph{Avg. Abs. Reduc.}: average absolute decrease in seconds (higher is better for both). Each miniF2F/PutnamBench/Putnam2025 proof is compiled five times. Reported values are means across proofs; subscripts denote standard deviations across the five compilations. ProofOptimizer is not evaluated on Putnam2025 (model weights not publicly released). Best per dataset in bold. 
% \textcolor{red}{variance is over multiple compilations, not multiple LLM generations.}
}
\label{tab:compilation_time_combined}
\resizebox{\linewidth}{!}{%
\begin{tabular}{llcc}
\toprule
\textbf{Dataset} & \textbf{Method} & \textbf{Avg. Rel. Reduc. $\uparrow$ (\%)} & \textbf{Avg. Abs. Reduc. $\uparrow$ (s)} \\
\midrule
\multirow{5}{*}{miniF2F~\cite{minif2f}}
 & ProofOptimizer~\cite{gu2025proofoptimizer}        & $-12.89_{\,\pm 0.55}$   & $0.97_{\,\pm 0.05}$ \\
 & Lean Refactor w/o Retrieval                       & $26.56_{\,\pm 0.38}$    & $3.02_{\,\pm 0.04}$ \\
 & Lean Refactor w/ Random Retrieval                 & $28.42_{\,\pm 0.48}$    & $3.11_{\,\pm 0.05}$ \\
 & Lean Refactor                                     & $-103.98_{\,\pm 2.64}$  & $1.46_{\,\pm 0.06}$ \\
 & \textbf{Lean Refactor w/ Reranking}               & $\mathbf{30.17}_{\,\pm 0.64}$ & $\mathbf{3.14}_{\,\pm 0.04}$ \\
\midrule
\multirow{5}{*}{PutnamBench~\cite{putnam}}
 & ProofOptimizer~\cite{gu2025proofoptimizer}        & $21.14_{\,\pm 0.43}$    & $8.16_{\,\pm 0.07}$ \\
 & Lean Refactor w/o Retrieval                       & $36.32_{\,\pm 0.18}$    & $7.05_{\,\pm 0.08}$ \\
 & Lean Refactor w/ Random Retrieval                 & $35.10_{\,\pm 0.35}$    & $6.77_{\,\pm 0.07}$ \\
 & Lean Refactor                                     & $37.11_{\,\pm 0.75}$    & $8.97_{\,\pm 0.11}$ \\
 & \textbf{Lean Refactor w/ Reranking}               & $\mathbf{42.62}_{\,\pm 0.27}$ & $\mathbf{10.01}_{\,\pm 0.10}$ \\
\midrule
\multirow{4}{*}{Putnam2025}
 & Lean Refactor w/o Retrieval                       & $49.53_{\,\pm 0.32}$    & $4.98_{\,\pm 0.08}$ \\
 & Lean Refactor w/ Random Retrieval                 & $50.13_{\,\pm 0.41}$    & $4.99_{\,\pm 0.08}$ \\
 & Lean Refactor                                     & $56.21_{\,\pm 0.34}$    & $5.58_{\,\pm 0.06}$ \\
 & \textbf{Lean Refactor w/ Reranking}               & $\mathbf{60.14}_{\,\pm 0.45}$ & $\mathbf{6.05}_{\,\pm 0.06}$ \\
\bottomrule
\end{tabular}
}
\vspace{-2em}
\end{table}

We redirect the framework from length to compilation time by swapping only the retrieval rule (Section~\ref{sec:multi_objective_retrieval}): \textbf{Lean Refactor w/ Reranking} reranks a semantically-similar candidate pool by each strategy's annotated compile-cost-reduction metadata. Controls are \textbf{w/o Retrieval}, \textbf{w/ Random Retrieval}, and the cosine-similarity-only \textbf{Lean Refactor}; ProofOptimizer is omitted on Putnam2025 (unreleased model weights). Compile time is measured via \texttt{lake env lean --profile}, excluding library import time. Table~\ref{tab:compilation_time_combined} reports average relative and absolute reductions on competition benchmarks; for research-level proofs, whose multi-file structure precludes clean per-proof wall-clock attribution, we instead measure heartbeats, Lean~4's deterministic proxy for elaboration work (Appendix~\ref{sec:compilation_time_research_level}).

\textbf{Lean Refactor w/ Reranking achieves the largest compilation-cost reduction on every competition and research-level benchmark} (see Table~\ref{tab:heartbeats_research} for research-level results), confirming that objective-aligned retrieval translates compile-cost metadata into wall-clock and heartbeat savings. miniF2F exposes the cost of objective \emph{misalignment}: cosine-only retrieval yields a positive absolute reduction but a strongly negative relative reduction ($-103.98\%$, i.e., compile time doubles on average in relative terms), driven by a few proofs whose compile time blows up by large factors, evidence that cosine-similarity-only retrieval can surface token-compressing yet compile-heavy strategies. Since ProofOptimizer's heartbeat-optimized proof variants are unreleased, we cannot measure their compilation time; therefore, Table~\ref{tab:compilation_time_combined} compares against their length-optimized proofs. However, under the heartbeat metric, \textbf{Lean Refactor w/ Reranking also beats ProofOptimizer's heartbeat-optimized numbers on miniF2F and PutnamBench} (Appendix~\ref{sec:heartbeat_comparison}). More broadly, these results confirm the framework's adaptability: unlike models fine-tuned for a single objective, it is redirected across objectives by adjusting retrieval criteria over densely annotated metadata, with no retraining required.

\subsection{Cross-Version Compatibility}
\label{sec:results_compatibility}
The rapid release cycles of Lean and Mathlib make toolchain drift a recurring concern. We test (i) whether version-tagged retrieval metadata enables refactoring against arbitrary target Lean versions and transfers across LLM backbones, and (ii) whether refactored proofs remain type-checkable as Mathlib evolves past their source version.

\vspace{-1em}
\begin{wraptable}{r}{0.55\linewidth}
\centering
\vspace{-1em}
\caption{\textbf{Cross-version compatibility on PutnamBench.} Average relative length reduction (\%, higher is better). Best per (version, model) in bold. Budget is 30.}
\label{tab:lean_compatibility}
\resizebox{\linewidth}{!}{%
\begin{tabular}{llccc}
\toprule
\textbf{Version} & \textbf{Model} & \textbf{No Retrieval} & \textbf{Full Retrieval} & \textbf{Filtered Retrieval} \\
\midrule
\multirow{2}{*}{v4.22.0} & Gemini 3 Flash & 55.99 & 55.43 & \textbf{57.36} \\
                         & GPT-OSS-20B        & 27.56 & 27.80 & \textbf{30.33} \\
\midrule
\multirow{2}{*}{v4.16.0} & Gemini 3 Flash & 60.61 & 60.41 & \textbf{62.24} \\
                         & GPT-OSS-20B        & 27.11 & 28.37 & \textbf{28.98} \\
\midrule
\multirow{2}{*}{v4.14.0} & Gemini 3 Flash & 58.51 & \textbf{62.85} & 59.54 \\
                         & GPT-OSS-20B       & \textbf{29.37} & 29.14 & 27.58 \\
\bottomrule
\end{tabular}
}
\vspace{-1em}
\end{wraptable}

\paragraph{Version-aware retrieval across Lean releases.}
We restrict cross-version evaluation to PutnamBench: research-level projects are pinned to specific Lean/Mathlib toolchains through deep intra-project dependencies, so changing versions breaks compilation regardless of refactoring quality, whereas PutnamBench proofs are self-contained. The lack of version-portable research benchmarks is a notable limitation for evaluating compatibility. We evaluate three Lean versions (v4.14.0, v4.16.0, v4.22.0) under three retrieval settings: \emph{No Retrieval}, \emph{Full Retrieval} (all strategies), and \emph{Filtered Retrieval} (strategies whose compatibility set covers the target version). Mathlib-derived strategies are excluded since Mathlib is locked to specific toolchains. We run both Gemini 3 Flash and GPT-OSS-20B; the latter is chosen because its knowledge cutoff predates \emph{every} version we test (Figure~\ref{fig:repo_compatibility}), making it a clean test of whether version-filtered retrieval can supply toolchain-specific knowledge the backbone lacks. Results are reported in Table~\ref{tab:lean_compatibility}.

Three findings emerge. \textbf{(1)} Filtered Retrieval attains the best length reduction on four of six (version, model) cells, and the pattern holds across both backbones. \textbf{(2)} Full Retrieval shows no consistent gain over No Retrieval, suggesting that unfiltered semantic neighbors can inject incompatible artifacts that disrupt refactoring. \textbf{(3)} On GPT-OSS-20B, Filtered Retrieval still beats both baselines at v4.22.0 and v4.16.0 despite the lack of pretraining exposure to those toolchains, providing direct evidence that the strategy bank compensates for missing in-weights knowledge. The exception is v4.14.0, where gains are less consistent, which we attribute to a significantly shrunken pool of strictly backward-compatible strategies.

\paragraph{Zero-shot version transfer of refactored proofs.}
\begin{wraptable}{r}{0.45\linewidth}
\centering
\vspace{-1.2em}
\caption{\textbf{Zero-shot version transfer on miniF2F.} Count of v4.19.0 proofs that still type-check at later releases without further refactoring. Best per row in bold. See Table~\ref{tab:lean_durability} for the full results.} 
\label{tab:lean_zero_shot_transfer}
\resizebox{\linewidth}{!}{%
\begin{tabular}{lccc}
\toprule
\textbf{Version} & \textbf{Lean Refactor} & \textbf{ProofOptimizer} & \textbf{Original} \\
\midrule
v4.20.1 & 194 & 194 & 194 \\
v4.22.0 & \textbf{189} & 188 & 187 \\
v4.26.0 & \textbf{168} & 162 & 141 \\
v4.29.0 & \textbf{166} & 161 & 140 \\
\bottomrule
\end{tabular}%
}
\vspace{-2em}
\end{wraptable}
We further evaluate zero-shot version transfer of refactored proofs by recompiling v4.19.0 outputs from Lean Refactor, ProofOptimizer, and the original verbose proofs against Mathlib releases through v4.29.0, without further refactoring. On miniF2F, Lean Refactor leads from v4.22.0 onward and the gap widens at later releases (Table~\ref{tab:lean_zero_shot_transfer}), showing that its refactored proofs continue to type-check across a longer span of future versions than both ProofOptimizer's proofs and the originals. On PutnamBench, Lean Refactor remains the strongest among proof optimization systems, though both shortened variants trail the originals, reflecting corpus-level differences in how proof length and version transfer interact. See Appendix~\ref{sec:durability} for full per-version results on both benchmarks.

\section{Related Work}
LLM-based provers \cite{ren2025deepseek, lin2025goedelproverv2scalingformaltheorem, chen2025seedprover15masteringundergraduatelevel} optimize correctness via RL but leave length, compile cost, and version compatibility unsupervised. Prior Lean refactoring \cite{ahuja2024improver, gu2025proofoptimizer} targets a single axis: ImProver needs per-objective example corpora, ProofOptimizer fine-tunes for length with compile cost demonstrated only as an alternative inference-time objective, and neither handles Lean/Mathlib version drift. General-purpose LLM code refactoring \cite{gautam2025refactorbench, gong2025language, karabiyik2025refactorgpt, bai2025polo, he2025swe, gong2025tuning} assumes paired data, stable APIs, and test-suite validity, conditions Lean inverts. Lean Refactor externalizes refactoring knowledge into a shared strategy bank with compile-time and version metadata, turning new objectives into retrieval rules rather than training runs, with a model-agnostic, training-free agent loop. Extended related work
is provided in Appendix~\ref{sec:extended_related_works}.

\section{Limitations}
\label{sec:limitations}
Cross-version evaluation is limited to PutnamBench due to toolchain-pinned research projects, and strategy-level metadata uses conservative cluster-level aggregation rather than marginal attribution; see Appendix~\ref{sec:limitations_extended} for details.

\section{Conclusion}
We present Lean Refactor, an LLM-agnostic framework that decouples Lean proof refactoring logic from the underlying model via a densely annotated strategy bank, steering frozen agentic LLMs to balance proof conciseness, compilation efficiency, and toolchain compatibility. Comprehensive evaluations across competition suites and research repositories show that metadata-aware retrieval is critical, and that the distilled strategies transfer across LLM backbones and even improve the general-purpose agent Claude Code. Ultimately, combining this retrieval mechanism with our specialized planner and compiler-in-the-loop architecture yields advanced refactoring capabilities that general-purpose coding agents struggle to match, laying the groundwork for more maintainable and version-resilient formal mathematical proofs.

\section*{Acknowledgment}
This research used resources of the National Energy Research Scientific Computing Center, a DOE Office of Science User Facility supported by the Office of Science of the U.S. Department of Energy under Contract No. DE-AC02-05CH11231 using NERSC award NERSC DDR-ERCAP0034682 and ASCR-ERCAP0031463.
ZW is supported by NSF Award 2523383 (DMS AIMING).

% \section*{References}
\bibliography{neurips_2026}

\begin{thebibliography}{10}

\bibitem{achiam2023gpt}
Josh Achiam, Steven Adler, Sandhini Agarwal, Lama Ahmad, Ilge Akkaya, Florencia~Leoni Aleman, Diogo Almeida, Janko Altenschmidt, Sam Altman, Shyamal Anadkat, et~al.
\newblock Gpt-4 technical report.
\newblock {\em arXiv preprint arXiv:2303.08774}, 2023.

\bibitem{achim2025aristotle}
Tudor Achim, Alex Best, Alberto Bietti, Kevin Der, Math{\"\i}s F{\'e}d{\'e}rico, Sergei Gukov, Daniel Halpern-Leistner, Kirsten Henningsgard, Yury Kudryashov, Alexander Meiburg, et~al.
\newblock Aristotle: Imo-level automated theorem proving.
\newblock {\em arXiv preprint arXiv:2510.01346}, 2025.

\bibitem{ahuja2024improver}
Riyaz Ahuja, Jeremy Avigad, Prasad Tetali, and Sean Welleck.
\newblock Improver: Agent-based automated proof optimization.
\newblock {\em arXiv preprint arXiv:2410.04753}, 2024.

\bibitem{alphaproof}
AlphaProof and AlphaGeometry teams.
\newblock {AI} achieves silver-medal standard solving international mathematical olympiad problems.
\newblock \url{https://deepmind.google/discover/blog/ai-solves-imo-problems-at-silver-medal-level/}, 2024.

\bibitem{anthropic2025claude45haiku}
{Anthropic}.
\newblock Claude 4.5 haiku model card, October 2025.
\newblock Accessed: 2026-04-22.

\bibitem{baanen2025growing}
Anne Baanen, Matthew~Robert Ballard, Johan Commelin, Bryan Gin-ge Chen, Michael Rothgang, and Damiano Testa.
\newblock Growing mathlib: maintenance of a large scale mathematical library.
\newblock In {\em International Conference on Intelligent Computer Mathematics}, pages 51--70. Springer, 2025.

\bibitem{bai2025polo}
Jiameng Bai, Ruoyi Xu, Sai Wu, Dingyu Yang, Junbo Zhao, and Gang Chen.
\newblock Polo: An llm-powered project-level code performance optimization framework.
\newblock In {\em Proceedings of the 34th International Joint Conference on Artificial Intelligence (IJCAI). IJCAI}, pages 7319--7328, 2025.

\bibitem{FLT_Lean}
Kevin Buzzard and Richard Taylor.
\newblock {FLT}.
\newblock \url{https://github.com/ImperialCollegeLondon/FLT}, 2025.

\bibitem{chen2025seedprover15masteringundergraduatelevel}
Jiangjie Chen, Wenxiang Chen, Jiacheng Du, Jinyi Hu, Zhicheng Jiang, Allan Jie, Xiaoran Jin, Xing Jin, Chenggang Li, Wenlei Shi, Zhihong Wang, Mingxuan Wang, Chenrui Wei, Shufa Wei, Huajian Xin, Fan Yang, Weihao Gao, Zheng Yuan, Tianyang Zhan, Zeyu Zheng, Tianxi Zhou, and Thomas~Hanwen Zhu.
\newblock Seed-prover 1.5: Mastering undergraduate-level theorem proving via learning from experience, 2025.

\bibitem{chen2025seed}
Luoxin Chen, Jinming Gu, Liankai Huang, Wenhao Huang, Zhicheng Jiang, Allan Jie, Xiaoran Jin, Xing Jin, Chenggang Li, Kaijing Ma, et~al.
\newblock Seed-prover: Deep and broad reasoning for automated theorem proving.
\newblock {\em arXiv preprint arXiv:2507.23726}, 2025.

\bibitem{cui2025large}
Bowen Cui, Tejas Ramesh, Oscar Hernandez, and Keren Zhou.
\newblock Do large language models understand performance optimization?
\newblock {\em arXiv preprint arXiv:2503.13772}, 2025.

\bibitem{de2015lean}
Leonardo De~Moura, Soonho Kong, Jeremy Avigad, Floris Van~Doorn, and Jakob von Raumer.
\newblock The lean theorem prover (system description).
\newblock In {\em Automated Deduction-CADE-25: 25th International Conference on Automated Deduction, Berlin, Germany, August 1-7, 2015, Proceedings 25}, pages 378--388. Springer, 2015.

\bibitem{depalma2024exploring}
Kayla DePalma, Izabel Miminoshvili, Chiara Henselder, Kate Moss, and Eman~Abdullah AlOmar.
\newblock Exploring chatgpt’s code refactoring capabilities: An empirical study.
\newblock {\em Expert Systems with Applications}, 249:123602, 2024.

\bibitem{gautam2025refactorbench}
Dhruv Gautam, Spandan Garg, Jinu Jang, Neel Sundaresan, and Roshanak~Zilouchian Moghaddam.
\newblock Refactorbench: Evaluating stateful reasoning in language agents through code.
\newblock {\em arXiv preprint arXiv:2503.07832}, 2025.

\bibitem{gong2025tuning}
Jingzhi Gong, Rafail Giavrimis, Paul Brookes, Vardan Voskanyan, Fan Wu, Mari Ashiga, Matthew Truscott, Mike Basios, Leslie Kanthan, Jie Xu, et~al.
\newblock Tuning llm-based code optimization via meta-prompting: An industrial perspective.
\newblock {\em arXiv preprint arXiv:2508.01443}, 2025.

\bibitem{gong2025language}
Jingzhi Gong, Vardan Voskanyan, Paul Brookes, Fan Wu, Wei Jie, Jie Xu, Rafail Giavrimis, Mike Basios, Leslie Kanthan, and Zheng Wang.
\newblock Language models for code optimization: Survey, challenges and future directions.
\newblock {\em arXiv preprint arXiv:2501.01277}, 2025.

\bibitem{gu2025proofoptimizer}
Alex Gu, Bartosz Piotrowski, Fabian Gloeckle, Kaiyu Yang, and Aram~H Markosyan.
\newblock Proofoptimizer: Training language models to simplify proofs without human demonstrations.
\newblock {\em arXiv preprint arXiv:2510.15700}, 2025.

\bibitem{he2025swe}
Xinyi He, Qian Liu, Mingzhe Du, Lin Yan, Zhijie Fan, Yiming Huang, Zejian Yuan, and Zejun Ma.
\newblock Swe-perf: Can language models optimize code performance on real-world repositories?
\newblock {\em arXiv preprint arXiv:2507.12415}, 2025.

\bibitem{karabiyik2025refactorgpt}
Muhammed~Abdulhamid Karabiyik.
\newblock Refactorgpt: a chatgpt-based multi-agent framework for automated code refactoring.
\newblock {\em PeerJ Computer Science}, 11:e3257, 2025.

\bibitem{li2024numinamath}
Jia Li, Edward Beeching, Lewis Tunstall, Ben Lipkin, Roman Soletskyi, Shengyi Huang, Kashif Rasul, Longhui Yu, Albert~Q. Jiang, Ziju Shen, Zihan Qin, Bin Dong, Li~Zhou, Yann Fleureau, Guillaume Lample, and Stanislas Polu.
\newblock Numinamath: The largest public dataset in ai4maths with 860k pairs of competition math problems and solutions.
\newblock \url{https://github.com/project-numina/aimo-progress-prize/blob/main/report/numina_dataset.pdf}, 2024.

\bibitem{lin2025goedel}
Yong Lin, Shange Tang, Bohan Lyu, Jiayun Wu, Hongzhou Lin, Kaiyu Yang, Jia Li, Mengzhou Xia, Danqi Chen, Sanjeev Arora, et~al.
\newblock Goedel-prover: A frontier model for open-source automated theorem proving.
\newblock {\em arXiv preprint arXiv:2502.07640}, 2025.

\bibitem{lin2025goedelproverv2scalingformaltheorem}
Yong Lin, Shange Tang, Bohan Lyu, Ziran Yang, Jui-Hui Chung, Haoyu Zhao, Lai Jiang, Yihan Geng, Jiawei Ge, Jingruo Sun, Jiayun Wu, Jiri Gesi, Ximing Lu, David Acuna, Kaiyu Yang, Hongzhou Lin, Yejin Choi, Danqi Chen, Sanjeev Arora, and Chi Jin.
\newblock Goedel-prover-v2: Scaling formal theorem proving with scaffolded data synthesis and self-correction, 2025.

\bibitem{liu2025atlasautoformalizingtheoremslifting}
Xiaoyang Liu, Kangjie Bao, Jiashuo Zhang, Yunqi Liu, Yu~Chen, Yuntian Liu, Yang Jiao, and Tao Luo.
\newblock Atlas: Autoformalizing theorems through lifting, augmentation, and synthesis of data, 2025.

\bibitem{lu2025lean}
Jialin Lu, Kye Emond, Kaiyu Yang, Swarat Chaudhuri, Weiran Sun, and Wuyang Chen.
\newblock Lean finder: Semantic search for mathlib that understands user intents.
\newblock {\em arXiv preprint arXiv:2510.15940}, 2025.

\bibitem{mathlib_breakage}
mathlib4.
\newblock Dealing with breakages from updating.
\newblock \url{https://github.com/leanprover-community/mathlib4/wiki/Using-mathlib4-as-a-dependency#dealing-with-breakages-from-updating}, 2025.
\newblock GitHub repository.

\bibitem{moura2021lean}
Leonardo~de Moura and Sebastian Ullrich.
\newblock The {Lean} 4 theorem prover and programming language.
\newblock 2021.

\bibitem{openai2025gptoss120bgptoss20bmodel}
OpenAI, :, Sandhini Agarwal, Lama Ahmad, Jason Ai, Sam Altman, Andy Applebaum, Edwin Arbus, Rahul~K. Arora, Yu~Bai, Bowen Baker, Haiming Bao, Boaz Barak, Ally Bennett, Tyler Bertao, Nivedita Brett, Eugene Brevdo, Greg Brockman, Sebastien Bubeck, Che Chang, Kai Chen, Mark Chen, Enoch Cheung, Aidan Clark, Dan Cook, Marat Dukhan, Casey Dvorak, Kevin Fives, Vlad Fomenko, Timur Garipov, Kristian Georgiev, Mia Glaese, Tarun Gogineni, Adam Goucher, Lukas Gross, Katia~Gil Guzman, John Hallman, Jackie Hehir, Johannes Heidecke, Alec Helyar, Haitang Hu, Romain Huet, Jacob Huh, Saachi Jain, Zach Johnson, Chris Koch, Irina Kofman, Dominik Kundel, Jason Kwon, Volodymyr Kyrylov, Elaine~Ya Le, Guillaume Leclerc, James~Park Lennon, Scott Lessans, Mario Lezcano-Casado, Yuanzhi Li, Zhuohan Li, Ji~Lin, Jordan Liss, Lily, Liu, Jiancheng Liu, Kevin Lu, Chris Lu, Zoran Martinovic, Lindsay McCallum, Josh McGrath, Scott McKinney, Aidan McLaughlin, Song Mei, Steve Mostovoy, Tong Mu, Gideon Myles, Alexander Neitz, Alex Nichol, Jakub
  Pachocki, Alex Paino, Dana Palmie, Ashley Pantuliano, Giambattista Parascandolo, Jongsoo Park, Leher Pathak, Carolina Paz, Ludovic Peran, Dmitry Pimenov, Michelle Pokrass, Elizabeth Proehl, Huida Qiu, Gaby Raila, Filippo Raso, Hongyu Ren, Kimmy Richardson, David Robinson, Bob Rotsted, Hadi Salman, Suvansh Sanjeev, Max Schwarzer, D.~Sculley, Harshit Sikchi, Kendal Simon, Karan Singhal, Yang Song, Dane Stuckey, Zhiqing Sun, Philippe Tillet, Sam Toizer, Foivos Tsimpourlas, Nikhil Vyas, Eric Wallace, Xin Wang, Miles Wang, Olivia Watkins, Kevin Weil, Amy Wendling, Kevin Whinnery, Cedric Whitney, Hannah Wong, Lin Yang, Yu~Yang, Michihiro Yasunaga, Kristen Ying, Wojciech Zaremba, Wenting Zhan, Cyril Zhang, Brian Zhang, Eddie Zhang, and Shengjia Zhao.
\newblock gpt-oss-120b \& gpt-oss-20b model card, 2025.

\bibitem{ospanov2025apollo}
Azim Ospanov, Farzan Farnia, and Roozbeh Yousefzadeh.
\newblock Apollo: Automated llm and lean collaboration for advanced formal reasoning.
\newblock {\em arXiv preprint arXiv:2505.05758}, 2025.

\bibitem{oueslati2025refagent}
Khouloud Oueslati, Maxime Lamothe, and Foutse Khomh.
\newblock Refagent: A multi-agent llm-based framework for automatic software refactoring.
\newblock {\em arXiv preprint arXiv:2511.03153}, 2025.

\bibitem{ouyang2022training}
Long Ouyang, Jeffrey Wu, Xu~Jiang, Diogo Almeida, Carroll Wainwright, Pamela Mishkin, Chong Zhang, Sandhini Agarwal, Katarina Slama, Alex Ray, et~al.
\newblock Training language models to follow instructions with human feedback.
\newblock {\em Advances in neural information processing systems}, 35:27730--27744, 2022.

\bibitem{peng2025criticleancriticguidedreinforcementlearning}
Zhongyuan Peng, Yifan Yao, Kaijing Ma, Shuyue Guo, Yizhe Li, Yichi Zhang, Chenchen Zhang, Yifan Zhang, Zhouliang Yu, Luming Li, Minghao Liu, Yihang Xia, Jiawei Shen, Yuchen Wu, Yixin Cao, Zhaoxiang Zhang, Wenhao Huang, Jiaheng Liu, and Ge~Zhang.
\newblock Criticlean: Critic-guided reinforcement learning for mathematical formalization, 2025.

\bibitem{piao2025refactoring}
Yonnel Chen~Kuang Piao, Jean~Carlors Paul, Leuson Da~Silva, Arghavan~Moradi Dakhel, Mohammad Hamdaqa, and Foutse Khomh.
\newblock Refactoring with llms: Bridging human expertise and machine understanding.
\newblock {\em arXiv preprint arXiv:2510.03914}, 2025.

\bibitem{polu2020generative}
Stanislas Polu and Ilya Sutskever.
\newblock Generative language modeling for automated theorem proving.
\newblock {\em arXiv preprint arXiv:2009.03393}, 2020.

\bibitem{ren2025deepseek}
ZZ~Ren, Zhihong Shao, Junxiao Song, Huajian Xin, Haocheng Wang, Wanjia Zhao, Liyue Zhang, Zhe Fu, Qihao Zhu, Dejian Yang, et~al.
\newblock Deepseek-prover-v2: Advancing formal mathematical reasoning via reinforcement learning for subgoal decomposition.
\newblock {\em arXiv preprint arXiv:2504.21801}, 2025.

\bibitem{Stehling2026Lean}
Rodrigo Stehling, Jialin Lu, Wuyang Chen, and Weiran Sun.
\newblock Lean formalization of pde topics.
\newblock \url{https://github.com/weiran-sun/pde}, 1 2026.
\newblock GitHub repository.

\bibitem{tao_lean_analysis}
Terence Tao.
\newblock A {Lean} companion to {Analysis I}.
\newblock \url{https://github.com/teorth/analysis}, 2024.

\bibitem{pfr_formalization}
Terence Tao.
\newblock Formalization of the {Polynomial Freiman-Ruzsa} conjecture of {Marton}.
\newblock \url{https://github.com/teorth/pfr}, 2025.

\bibitem{geminiteam2025gemini3flash}
{The Gemini Team}.
\newblock Gemini 3 flash: frontier intelligence built for speed.
\newblock \url{https://blog.google/products-and-platforms/products/gemini/gemini-3-flash/}, December 2025.
\newblock Accessed: 2026-03-16.

\bibitem{physlib_lean}
Joseph Tooby-Smith.
\newblock {physlib}: A project to digitalise results from physics into {Lean}.
\newblock \url{https://github.com/leanprover-community/physlib}, 2025.

\bibitem{trinh2024solving}
Trieu~H Trinh, Yuhuai Wu, Quoc~V Le, He~He, and Thang Luong.
\newblock Solving olympiad geometry without human demonstrations.
\newblock {\em Nature}, 625(7995):476--482, 2024.

\bibitem{putnam}
George Tsoukalas, Jasper Lee, John Jennings, Jimmy Xin, Michelle Ding, Michael Jennings, Amitayush Thakur, and Swarat Chaudhuri.
\newblock Putnambench: Evaluating neural theorem-provers on the putnam mathematical competition, 2024.

\bibitem{wu2025fasterpy}
Yue Wu, Minghao Han, Ruiyin Li, Peng Liang, Amjed Tahir, Zengyang Li, Qiong Feng, and Mojtaba Shahin.
\newblock Fasterpy: An llm-based code execution efficiency optimization framework.
\newblock {\em arXiv preprint arXiv:2512.22827}, 2025.

\bibitem{xin2024deepseek}
Huajian Xin, Daya Guo, Zhihong Shao, Zhizhou Ren, Qihao Zhu, Bo~Liu, Chong Ruan, Wenda Li, and Xiaodan Liang.
\newblock Deepseek-prover: Advancing theorem proving in llms through large-scale synthetic data.
\newblock {\em arXiv preprint arXiv:2405.14333}, 2024.

\bibitem{xin2024deepseekv15}
Huajian Xin, ZZ~Ren, Junxiao Song, Zhihong Shao, Wanjia Zhao, Haocheng Wang, Bo~Liu, Liyue Zhang, Xuan Lu, Qiushi Du, et~al.
\newblock Deepseek-prover-v1.5: Harnessing proof assistant feedback for reinforcement learning and monte-carlo tree search.
\newblock {\em arXiv preprint arXiv:2408.08152}, 2024.

\bibitem{yang2025perfcoder}
Jiuding Yang, Shengyao Lu, Hongxuan Liu, Shayan Shirahmad~Gale Bagi, Zahra Fazel, Tomasz Czajkowski, and Di~Niu.
\newblock Perfcoder: Large language models for interpretable code performance optimization.
\newblock {\em arXiv preprint arXiv:2512.14018}, 2025.

\bibitem{yang2023leandojo}
Kaiyu Yang, Aidan Swope, Alex Gu, Rahul Chalamala, Peiyang Song, Shixing Yu, Saad Godil, Ryan~J Prenger, and Animashree Anandkumar.
\newblock Leandojo: Theorem proving with retrieval-augmented language models.
\newblock {\em Advances in Neural Information Processing Systems}, 36:21573--21612, 2023.

\bibitem{ye2026verinabenchmarkingverifiablecode}
Zhe Ye, Zhengxu Yan, Jingxuan He, Timothe Kasriel, Kaiyu Yang, and Dawn Song.
\newblock Verina: Benchmarking verifiable code generation, 2026.

\bibitem{zhang2025qwen3embeddingadvancingtext}
Yanzhao Zhang, Mingxin Li, Dingkun Long, Xin Zhang, Huan Lin, Baosong Yang, Pengjun Xie, An~Yang, Dayiheng Liu, Junyang Lin, Fei Huang, and Jingren Zhou.
\newblock Qwen3 embedding: Advancing text embedding and reranking through foundation models, 2025.

\bibitem{zhao2025semopt}
Yuwei Zhao, Yuan-An Xiao, Qianyu Xiao, Zhao Zhang, and Yingfei Xiong.
\newblock Semopt: Llm-driven code optimization via rule-based analysis.
\newblock {\em arXiv preprint arXiv:2510.16384}, 2025.

\bibitem{minif2f}
Kunhao Zheng, Jesse~Michael Han, and Stanislas Polu.
\newblock Minif2f: a cross-system benchmark for formal olympiad-level mathematics, 2022.

\end{thebibliography}
\bibliographystyle{plain}

%%%%%%%%%%%%%%%%%%%%%%%%%%%%%%%%%%%%%%%%%%%%%%%%%%%%%%%%%%%%
\newpage
\appendix

\section{Extended Related Work}
\label{sec:extended_related_works}

\paragraph{LLM-based theorem proving in Lean.}
A rapid line of work trains LLMs to generate Lean proofs end-to-end with reinforcement learning from compiler-verified rewards, including DeepSeek-Prover \cite{xin2024deepseek, xin2024deepseekv15, ren2025deepseek}, Goedel-Prover \cite{lin2025goedel, lin2025goedelproverv2scalingformaltheorem}, Seed-Prover \cite{chen2025seed, chen2025seedprover15masteringundergraduatelevel}, building on earlier tactic-level and retrieval-augmented systems \cite{polu2020generative, yang2023leandojo}. These rewards, however, primarily target proof correctness, leaving length, compilation cost, and version compatibility unsupervised. The resulting proofs are correct but verbose, slow to compile, and brittle across Mathlib releases \cite{ahuja2024improver, gu2025proofoptimizer}. Our framework operates downstream of these provers, refactoring their outputs under multiple cost objectives without retraining the underlying model.

\paragraph{Lean proof optimization.}
Two prior systems directly target Lean 4 proof simplification. ImProver \cite{ahuja2024improver} wraps a black-box LLM in an agentic loop combining Chain-of-States prompting, best-of-n sampling with iterative refinement, and retrieval over Lean/Mathlib documentation and per-metric example databases. ProofOptimizer \cite{gu2025proofoptimizer} instead fine-tunes a dedicated model on synthesized long-short pairs to directly shorten proofs. While both achieve length reductions, each restricts refactoring to a single-objective paradigm. ProofOptimizer's objective is frozen at training time. ImProver accommodates arbitrary user-defined metrics by using them to drive both example retrieval and candidate selection; however, scaling to new objectives requires constructing an entirely new, metric-specific example corpus. Crucially, neither system addresses the broader bottlenecks of Lean 4 proof maintenance identified in Section~\ref{sec:refactoring_strategies}. Compilation cost is treated as a secondary objective swapped in at inference rather than jointly optimized~\cite{gu2025proofoptimizer}, and continuous Lean 4 and Mathlib version drift remains invisible to both frozen fine-tuned models and standard example-based retrievers. Lean Refactor overcomes these limitations by retrieving \emph{refactoring strategies} densely annotated with multi-objective execution metadata, such as expected compilation-time reductions and validated Lean 4 version compatibility. This enables Lean Refactor to turn complex, competing objectives into configurable retrieval rules over a single, shared strategy bank, bypassing the need for training-time commitments or fractured, per-metric datasets.

\paragraph{LLM-based code refactoring and performance optimization.}
A broader literature studies LLM-driven refactoring \cite{gautam2025refactorbench, gong2025language} and code performance optimization for general-purpose languages. Agentic and prompting-based systems target maintainability and code quality \cite{depalma2024exploring, piao2025refactoring, oueslati2025refagent, karabiyik2025refactorgpt}, while a parallel thread pursues runtime speedup \cite{bai2025polo, he2025swe, cui2025large, zhao2025semopt, wu2025fasterpy, yang2025perfcoder, gong2025tuning}. These methods operate on settings such as Python, C++, and Java, where paired refactoring data is plentiful, APIs are comparatively stable, and correctness is verified by test suites rather than a strict type checker. Lean inverts each of these conditions: paired data is scarce, the toolchain breaks compilation weekly, and validity is a hard binary signal. Lean Refactor's agentic system is distinguished by the densely annotated strategy bank and a compiler-in-the-loop validation tailored to Lean's release dynamics.

\section{Details about Strategy Bank Construction}
\label{sec:strategy_bank_details}

This section provides a complete description of the data pipeline summarized in
Section~\ref{sec:strategy_bank}. We detail (i) the source repositories from
which formal statements are aggregated, (ii) the dual-strategy procedure used to synthesize
aligned long--short proof pairs, (iii) the profiling and cross-version recompilation pipeline that
produces the refactoring metadata, (iv) the prompting protocol used to distill location-grounded
refactoring strategies from each pair, (v) the quality-control filter, (vi) the iterative
embedding-based deduplication pipeline, and (vii) the procedure by which pair-level metadata is
aggregated into strategy-level annotations.
 
\subsection{Source Repositories of Lean Statements}
\label{app:sb-sources}
 
To ensure that our strategy bank generalizes across both competition mathematics and a broad
spectrum of undergraduate- and research-level domains, we aggregate formal statements from four
complementary repositories.
 
\begin{itemize}[leftmargin=*]
    \item \textbf{NuminaMath-1.5~\cite{li2024numinamath}.} A large-scale corpus of formalized
    competition problems. NuminaMath provides a high density of challenging Olympiad-style Lean
    statements that exercise the full range of tactical machinery typically employed in contest
    proofs. 
    \item \textbf{FineLeanCorpus~\cite{peng2025criticleancriticguidedreinforcementlearning}.}
    A second large-scale source of competition-level statements, complementary to NuminaMath in
    style and topical coverage.
    \item \textbf{Mathlib4.} The standard library for the Lean~4 ecosystem. We extract theorems
    from \texttt{Mathlib4} to represent broader mathematical disciplines that are underrepresented in
    competition-oriented corpora.
    \item \textbf{ATLAS~\cite{liu2025atlasautoformalizingtheoremslifting}.} A synthetic dataset
    of theorem statements lifted from \texttt{Mathlib4} contexts. ATLAS broadens the stylistic
    distribution of statements beyond what is naturally present in human-curated corpora and
    serves as a useful augmentation for diversity.
\end{itemize}
 
This hybrid sourcing strategy exposes the downstream pipeline to a wide range of Lean coding
styles, tactic preferences, and proof-architecture patterns. The held-out
evaluation set used to assess generalization is described separately in the experiments section
of the main paper.

\paragraph{Decontamination against the held-out evaluation set.} To preclude leakage from our held-out evaluation set into the long--short proof pairs used to build the strategy bank, we apply a two-stage decontamination filter to every theorem statement before proof synthesis. First, we perform exact-match deduplication on normalized statements. Second, we perform embedding-based near-duplicate detection: each candidate training statement is encoded with Qwen3-Embedding-8B~\cite{zhang2025qwen3embeddingadvancingtext} and compared against the embeddings of all evaluation statements. If the top-1 cosine similarity is below $0.8$, we treat the candidate as non-duplicate; if it is at or above $0.8$, we invoke GPT-OSS-120B as an LLM-as-judge with the candidate and the matched evaluation statement to make the final determination, and discard the candidate if the judge confirms duplication. This filter is applied upstream of proof synthesis and strategy distillation, so the resulting 200K long--short pairs, the 481{,}567 raw strategy extractions, and the final corpus of 9{,}237 unique strategies are all derived exclusively from theorems that survive both stages.
 
\subsection{Proof Synthesis and Long--Short Pair Construction}
\label{app:sb-pairs}
 
\paragraph{Proof generation.} With the exception of \texttt{Mathlib4}, the sourced repositories
provide only theorem statements without ground-truth proofs. We therefore synthesize formal
proofs using two prover models:
Goedel-Prover-V2-32B~\cite{lin2025goedelproverv2scalingformaltheorem} and
GPT-OSS-120B~\cite{openai2025gptoss120bgptoss20bmodel}. Using two provers rather than one is a
deliberate design choice: it broadens the distribution of tactic patterns, naming conventions,
and structural choices that subsequently appear in the long--short pairs and, by extension, in
the distilled strategies.
 
For each theorem statement we sample between one and three candidate proofs per model, with the
exact count bounded by our compute budget. To guarantee environmental consistency during
verification, every candidate is prepended with a standardized header containing the necessary
\texttt{import} directives and \texttt{namespace} declarations before being submitted to the
Lean~4 compiler. When multiple candidate proofs for the same theorem successfully verify, we
retain the \emph{longest} valid proof, since verbose proofs provide
the richest foundation for refactoring and yield the most informative long--short pairs in the
subsequent stage.
 
\paragraph{Aligned long--short pair construction.} Given a corpus of verified proofs, we
construct aligned pairs through a length-conditioned, dual-strategy prompting pipeline:
\begin{itemize}[leftmargin=*]
    \item \textbf{Forward simplification (long $\to$ short).} For proofs exceeding 50 tokens,
    we prompt GPT-OSS-120B to optimize the trajectory and produce a shorter counterpart.
    \item \textbf{Reverse complexification (short $\to$ long).} For proofs under 50 tokens, we
    prompt GPT-OSS-120B to synthetically expand the proof into a more verbose, less idiomatic
    counterpart, yielding the ``long'' element of the pair.
\end{itemize}
The reverse-complexification branch ensures that compact, idiomatic proofs, which would
otherwise be excluded because no naturally occurring ``long'' counterpart exists, still
contribute to the bank by surfacing the simplifications they implicitly embody.
 
To improve structural fidelity in both directions, we augment the prompts with explicit
dependency information that enumerates the lemmas and premises invoked in the input proof.
Empirically, this dependency context reduces hallucinated lemma references and produces
transformations that are more faithful to the underlying proof structure. Each candidate
transformed proof is recompiled to confirm that the produced pair consists of two
independently verified Lean proofs of the same theorem.
 
In total this pipeline yields \textbf{200K verified long--short proof pairs}, which form the foundation for all subsequent strategy distillation.
 
\subsection{Refactoring Metadata Annotation}
\label{app:sb-metadata}
 
Each long--short pair is annotated along two axes that the LLM cannot infer from its training
data alone: compilation-time reduction and Lean-toolchain compatibility.
 
\paragraph{Compilation-time profiling.} We measure the compilation time of both the long and
the short proof in each pair using the \texttt{lake env lean --profile} command, which exposes
fine-grained per-declaration timing. We explicitly isolate proof execution from library import
overhead, since import time is dominated by caching effects unrelated to the
proof content itself. Each measurement is taken in a consistent environment to ensure stable measurements, and we record the relative reduction in execution time achieved by the
shortened proof. This per-pair speedup is the unit from which strategy-level estimates are
later aggregated (Appendix~\ref{app:sb-aggregation}).
 
\paragraph{Cross-version compatibility testing.} To assess whether a refactoring remains valid
under future Lean releases, we recompile each shortened proof under multiple Lean and
\texttt{Mathlib} toolchains beyond the original v4.24.0: \texttt{v4.14.0}, \texttt{v4.16.0},
and \texttt{v4.22.0}. We restrict version-compatibility analysis to non-\texttt{Mathlib}
theorems, because \texttt{Mathlib} itself is tightly coupled to a specific Lean release via
\texttt{lean-toolchain} and routinely fails to build under manual version overrides; this
coupling makes any cross-version signal collected on \texttt{Mathlib} theorems an artifact of
infrastructure rather than of the refactoring strategy under study.
 
For each eligible pair we track the compilation status of both the long and short proof under
each toolchain. A refactored proof is deemed compatible with a target version if the short
proof compiles successfully in that version. The strategy is considered robust on the interval
spanned by the toolchains under which all of its shortened proofs continue to compile, a
notion that is formalized at the strategy level in Appendix~\ref{app:sb-aggregation}.
 
\subsection{Strategy Summarization with Location Grounding}
\label{app:sb-summarization}
 
A single long--short proof pair typically encompasses several distinct refactoring operations
applied to different regions of the code. To capture this granularity, we prompt
GPT-OSS-120B with both proofs and ask it to extract a set of applied refactoring strategies.
Crucially, we instruct the model to ground each extracted strategy to a specific component of
the original long proof, identified by start and end line numbers.
 
This grounding requirement serves two purposes:
\begin{enumerate}[leftmargin=*]
    \item \emph{Hallucination control.} In preliminary experiments we observed that without an
    explicit anchoring constraint, the model frequently fabricated strategies that did not
    correspond to any concrete change between the long and short proofs. Forcing each strategy
    to point to a specific code region eliminates a large class of these hallucinations and improves extraction quality.
    \item \emph{Retrieval supervision.} The proof-component-to-strategy mapping is the
    supervision signal used to fine-tune the strategy-retrieval model employed by our agent at
    inference time. Given a proof component currently being refactored, the retriever returns
    the strategies historically associated with similar components, providing contextually
    relevant guidance.
\end{enumerate}
 
Importantly, location grounding does not bias the extracted strategies toward purely localized
fixes. Global structural refactorings, such as restructuring an entire case split or replacing
a multi-step induction with a single library lemma, are accommodated by grounding the strategy
to a correspondingly larger contiguous chunk of the proof.
 
\subsection{Strategy Schema}
\label{app:sb-schema}
 
To ensure clarity, applicability, and reusability, each distilled strategy is organized
into the following fields. Together they form a self-contained record sufficient for the
agent to reason about \emph{when}, \emph{why}, and \emph{how} to apply the refactoring at
inference time. Examples are provided in
Appendix~\ref{sec:strategy_examples}.
 
\begin{itemize}[leftmargin=*]
    \item \textbf{Title.} A short natural-language summary of the high-level idea of the refactoring strategy.
    \item \textbf{Description.} A natural-language statement of the core conceptual idea
    behind the refactoring. It includes information like the semantic shift that the strategy brings (e.g.\ replacing a manual
    step-by-step existence construction or a redundant case analysis with a more powerful
    library lemma or automated tactic).
    \item \textbf{When to Apply.} A generalized description of the Lean code pattern or
    logical situation that serves as a precondition for the strategy. This field explicitly
    identifies the structural ``anti-patterns'' that warrant the proposed refactoring and
    enables the agent to recognize applicable contexts in unseen proofs.
    \item \textbf{Application Guide.} A step-by-step, actionable refactoring guide. It provides the
    agent with deterministic, sequential instructions for executing the transformation,
    minimizing ambiguity at the point of application.
    \item \textbf{Abstract Example.} A generalized exemplar that supplies a concrete
    ``before'' and ``after'' snippet of the transformation. Specific variable names and
    domain artifacts are abstracted away, so that the example highlights the underlying
    syntactic and tactical shift rather than the surface details of any one theorem.
    \item \textbf{Potential Reduction.} A categorical estimate (\textit{high}, \textit{medium},
    or \textit{low}) of the expected impact on proof length.
\end{itemize}

\subsection{Quality Filtering via LLM-as-Judge}
\label{app:sb-filtering}
 
To guarantee both the correctness of the distilled strategies and the fidelity of their
location grounding, we apply a rigorous filtering stage. Each extracted strategy is evaluated
by an LLM-as-judge alongside its corresponding long--short proof pair and the identified proof
component. The judge assesses two criteria:
\begin{enumerate}[leftmargin=*]
    \item \emph{Transformation correctness.} Whether the proposed strategy correctly and
    logically transforms the identified component of the long proof into its optimized
    counterpart in the short proof.
    \item \emph{Schema fidelity.} Whether all six fields of the strategy schema
    (Appendix~\ref{app:sb-schema}) are accurately and consistently populated, with no
    contradictions between, e.g., the \emph{When to Apply} precondition and the
    \emph{Abstract Example}.
\end{enumerate}
Only strategies passing both checks are retained for the next stage. Strategies that fail
either check are discarded; we do not attempt to repair them, as our preliminary experiments
indicated that repair attempts frequently introduce subtler errors that propagate downstream.
 
\subsection{Clustering and Iterative Deduplication}
\label{app:sb-deduplication}
 
The filtered set still contains substantial redundancy: many distinct long--short pairs surface
the same underlying refactoring idea, expressed with surface-level variation. To prevent
redundant strategies from consuming valuable prompt context at inference time, and to keep
retrieval efficient, we curate a compact, semantics-aware subset of unique strategies via the
following iterative pipeline.
 
\paragraph{Initialization.} We seed the unique-strategy set by prompting GPT-OSS-120B to identify a diverse initial collection of canonical strategies from a sampled subset of the pool. This produces a high-coverage starting point against which subsequent strategies can be compared.

\paragraph{Iterative deduplication.} Each remaining strategy is then processed sequentially against the dynamically growing unique set via a four-stage cascade:
\begin{enumerate}[leftmargin=*]
    \item \emph{Embedding and candidate retrieval.} We encode the strategy's \emph{Description} field with the Qwen3-Embedding-8B~\cite{zhang2025qwen3embeddingadvancingtext} model and retrieve the top-10 most semantically similar entries from the current unique set by cosine similarity.
    \item \emph{High-confidence shortcut.} If the top-1 cosine similarity is $\geq 0.9$, the strategy is treated as a duplicate of the matched entry without further inspection. We adopt this shortcut because manual inspection of borderline cases indicated that pairs above this threshold are essentially always paraphrases of the same underlying refactoring, and skipping the judge call at this end of the distribution yields a substantial reduction in pipeline cost without a measurable loss in cluster purity.
    \item \emph{Judge-based duplicate detection.} If the top-1 cosine similarity is $< 0.9$, we prompt GPT-OSS-120B as an LLM-as-judge with the current strategy together with the 10 retrieved candidates. The judge decides whether the current strategy is a semantic duplicate of any existing entry. This stage handles the long tail of paraphrastic variation that pure embedding similarity tends to underestimate (e.g., strategies that share an underlying tactical idea but use different terminology or describe it at different levels of abstraction).
    \item \emph{Update.} If either the shortcut or the judge identifies a duplicate, the current strategy is \emph{clustered} under the matched existing entry, contributing its pair-level metadata to that cluster's aggregate (Appendix~\ref{app:sb-aggregation}). Otherwise, the strategy is appended to the unique set as a new canonical entry and becomes a candidate for matching against subsequent strategies.
\end{enumerate}

By combining embedding-based shortlisting, a high-similarity shortcut, and an LLM judge fallback, this cascade keeps the procedure tractable on the full filtered pool while still benefiting from the judge's ability to resolve paraphrastic variation.

\paragraph{Final corpus.} Applying this pipeline to the filtered pool of 481{,}567 raw extractions yields a refined corpus of \textbf{9{,}237 unique refactoring strategies}, each backed by a cluster of constituent long--short pairs that share its semantics.
 
\subsection{Strategy-Level Metadata Aggregation}
\label{app:sb-aggregation}
 
The pair-level metadata collected in Appendix~\ref{app:sb-metadata} is lifted to the strategy
level by aggregating across the cluster of long--short pairs assigned to each unique strategy.
 
\paragraph{Compilation-time reduction.} For each strategy, we keep the median relative
execution-time reduction across all long--short pairs in its cluster to get a robust strategy-level estimate.
 
\paragraph{Version compatibility set.} Following Appendix~\ref{app:sb-metadata}, we restrict
attention to non-\texttt{Mathlib} pairs. A strategy's \emph{compatibility set} is defined as
the intersection of toolchains under which the shortened proofs of all its constituent pairs
compile successfully. This intersection-based definition is conservative by design: a strategy
is certified for a toolchain only if every observed instantiation of the strategy survives
recompilation under that toolchain. Aggregating across all constituent pairs, rather than
relying on a single example, ensures that a toolchain enters the compatibility set only when
the strategy has been shown to transfer reliably across diverse instantiations, rather than
happening to compile in one isolated case.

\paragraph{Downstream use.} The aggregated metadata is the lever that makes the bank
\emph{version-aware} and \emph{efficiency-aware} at inference time: the multi-objective
retrieval procedure (Section ~\ref{sec:multi_objective_retrieval}) uses the compatibility set to filter
strategies incompatible with the user's target toolchain and the median speedup to rank
surviving candidates. This is the mechanism by which we decouple refactoring reliability from
the LLM's training cutoff.

\section{Limitations}
\label{sec:limitations_extended}

\paragraph{Benchmark coverage for cross-version evaluation.}
Our cross-version compatibility experiments (Section~\ref{sec:results_compatibility}) are confined to PutnamBench because, to our knowledge, no existing Lean benchmark supports compatibility evaluation beyond self-contained competition problems. Research-level projects pin to specific Lean/Mathlib toolchains through deep intra-project dependencies, so changing the toolchain breaks compilation regardless of refactoring quality. Constructing version-portable research-level benchmarks, e.g., proofs whose dependencies are at a granularity that decouples them from the project's pinned toolchain, is an important direction for the community.

\paragraph{Joint-effect attribution.}
Strategy-level compile-time annotations are computed as the cluster median of pair-level reductions, where each pair contributes the joint speedup of its refactorings; analogously, version-compatibility sets are intersected across constituent pairs, so a strategy inherits the toolchain coverage that all its instantiations jointly satisfy. Both choices fit the role the metadata plays at inference: the retrieval module reranks candidates by \emph{relative} compile-time benefit and filters by guaranteed version coverage, for which the median and intersection are robust, conservative estimators. Disentangling marginal per-strategy contributions, via Shapley-style attribution, or regression over co-occurrence indicators, would further enrich the schema along a new axis and is a natural extension we leave to future work.

\section{An Example of Proof Length and Compilation Time Tradeoff}
\label{sec:len_vs_time}

To illustrate the misalignment of proof length and compilation time, consider the two alternative Lean proofs for the theorem \texttt{amc12\_2001\_p21} from miniF2F~\cite{minif2f} below.

\begin{minipage}[t]{0.5\textwidth}
\begin{lstlisting}[language=Lean]
theorem amc12_2001_p21
  (a b c d : N)
  (h0 : a * b * c * d = Nat.factorial 8)
  (h1 : a * b + a + b = 524)
  (h2 : b * c + b + c = 146)
  (h3 : c * d + c + d = 104) :
  a - d = (10 : Z) := by
  norm_num [Nat.factorial] at h0
  have : b <= 525 := by nlinarith
  interval_cases b <;> simp_all
  <;> nlinarith
\end{lstlisting}
\end{minipage}
\hfill
\begin{minipage}[t]{0.5\textwidth}
\begin{lstlisting}[language=Lean]
theorem amc12_2001_p21
  (a b c d : N)
  (h0 : a * b * c * d = Nat.factorial 8)
  (h1 : a * b + a + b = 524)
  (h2 : b * c + b + c = 146)
  (h3 : c * d + c + d = 104) :
  a - d = (10 : Z) := by
  -- Factor hypotheses into (x+1)(y+1) = xy + x + y + 1
  have h4 : (a+1)*(b+1) = 525 := by 
    -- proof omitted
  have h5 : (b+1)*(c+1) = 147 := by 
    -- proof omitted
  have h6 : (c+1)*(d+1) = 105 := by
    -- proof omitted
  -- Pin down each variable via gcd + interval_cases
  have h7  : b = 20 := by -- proof omitted
  have h8  : a = 24 := by -- proof omitted
  have h9  : c = 6  := by -- proof omitted
  have h10 : d = 14 := by -- proof omitted
  -- Cast to integers and conclude
  have h11 : (a:Z) - (d:Z) = 10 := by 
    -- proof omitted
  exact_mod_cast h11
\end{lstlisting}
\end{minipage}

The short proof is concise but relies on a single heavy cascade \texttt{interval\_cases, simp\_all, nlinarith} over the loose bound $b \leq 525$, so Lean's most expensive tactics fire on hundreds of branches. Conversely, the second proof spans over 130 lines of explicit, step-by-step arithmetic deductions. Despite its much longer text, it significantly reduces the compiler's search burden and compiles more than $20\times$ faster than its shorter counterpart. This stark contrast highlights that optimizing LLMs solely for shorter text lengths can drastically penalize underlying runtime metrics.

\section{Retrieval Model Training Details}
\label{sec:retrieval_model_training_details}
We initialize our strategy search model using Qwen3-Embedding-8B~\cite{zhang2025qwen3embeddingadvancingtext}. We fine-tune the model with a contrastive learning objective, using in-batch negative sampling and a margin-based false-negative filtering mechanism. We train the model for 1 epoch, a per-device batch size of 16, gradient accumulation steps of 2, and 4 NVIDIA 6000Ada GPUs. We use the AdamW optimizer with a learning rate of $6 \times 10^{-6}$ and $\epsilon = 1 \times 10^{-8}$. The learning rate is linearly warmed up over the first 300 steps, then decayed to 0 using a cosine schedule. Temperature $\tau$ is set to 0.01, and the margin $m$ is set to 0.1.

\section{Proof Dependency Reduction}
\label{sec:results_dependencies}

An essential measure of formal proof elegance and maintainability is a minimal reliance on external lemmas. Analyzing the length-optimized proofs from Section~\ref{sec:results_proof_length}, we find that Lean Refactor consistently reduces dependencies, defined as statements invoked from external libraries (e.g., Mathlib) or distinct intra-project modules. As shown in Tables~\ref{tab:dependency_reduction_competition} and \ref{tab:dependency_reduction_research}, dependencies decreased in 64.0\% of PutnamBench and 51.0\% of miniF2F proofs. Remarkably, this trend holds even in complex research-level mathematics typically entangled with extensive library architectures, yielding reductions in 66.7\% of the Analysis and 60.0\% of the FLT datasets. Across all benchmarks, cases where the agent increases dependencies remain rare.

This systematic reduction reveals that Lean Refactor achieves brevity through genuine logical simplification rather than merely offloading complexity to heavy, specialized library calls. By discovering direct pathways from local premises, the agent naturally produces highly self-contained proofs. This structural shift provides a significant practical advantage by eliminating the burden of memorizing or invoking vast numbers of external lemma names. Consequently, these optimized proofs are significantly more readable for human mathematicians, easier to maintain, and inherently more resilient to the cross-version library breakages discussed in Section~\ref{sec:results_compatibility}.

\vspace{-1em}
\begin{table}[H]
\centering
\caption{Comparison of Proof Dependency Counts across miniF2F, PutnamBench, and Putnam 2025 datasets. Reduced Dependencies indicate a reduction in the number of external (e.g., Mathlib) or intra-project dependencies required for the proof.}
\label{tab:dependency_reduction_competition}
\begin{tabular}{lccc}
\toprule
\textbf{Comparison Metric} & \textbf{miniF2F~\cite{minif2f}} & \textbf{PutnamBench~\cite{putnam}} & \textbf{Putnam 2025} \\
\midrule
Total Theorems Compared & 194 & 75 & 66 \\
Reduced Dependencies & 99 (51.0\%) & 48 (64.0\%) & 29 (43.9\%) \\
Increased Dependencies & 11 (5.7\%) & 11 (14.7\%) & 12 (18.2\%) \\
No Change & 84 (43.3\%) & 16 (21.3\%) & 25 (37.9\%) \\
\bottomrule
\end{tabular}
\end{table}

% term-based vs. tactic-based,

% the number of advanced tactics used, different used tactics;

\vspace{-1em}
\begin{table}[H]
\centering
\caption{Comparison of Proof Dependency Counts across Analysis, FLT, PFR, and Physlean datasets. Reduced Dependencies indicate a reduction in the number of external (e.g., Mathlib) or intra-project dependencies required for the proof.}
\label{tab:dependency_reduction_research}
\begin{tabular}{lcccc}
\toprule
\textbf{Comparison Metric} & \textbf{Analysis~\cite{tao_lean_analysis}} & \textbf{FLT~\cite{FLT_Lean}} & \textbf{PFR~\cite{pfr_formalization}} & \textbf{Physlean~\cite{physlib_lean}} \\
\midrule
Total Theorems Compared & 45 & 45 & 45 & 45 \\
Reduced Dependencies & 30 (66.7\%) & 27 (60.0\%) & 24 (53.3\%) & 19 (42.2\%) \\
Increased Dependencies & 8 (17.7\%) & 11 (24.4\%) & 7 (15.6\%) & 11 (24.5\%) \\
No Change & 7 (15.6\%) & 7 (15.6\%) & 14 (31.1\%) & 15 (33.3\%) \\
\bottomrule
\end{tabular}
\end{table}

\section{Zero-shot Version Transfer Details}
\label{sec:durability}
We complement the retrieval-side analysis with a passive audit of zero-shot version transfer: we take the proofs produced on a single source version (v4.19.0) and compile them against every subsequent Mathlib release through v4.29.0, recording how many still type-check without further refactoring. We compare our system (LeanRefactor) against the original human-written proofs and against ProofOptimizer, a shortening baseline run from the same source version. This probes whether the act of refactoring itself alters how well a proof transfers across Lean/Mathlib releases, independent of any retrieval-time filtering. Table~\ref{tab:lean_durability} reports the results.

\begin{table}[ht]
\centering
\caption{\textbf{Zero-shot version transfer of refactored proofs.} Number of v4.19.0 proofs from each system that still type-check at each later Mathlib release without further refactoring. Totals: 75 (PutnamBench), 194 (MiniF2F).}
\label{tab:lean_durability}
\resizebox{\linewidth}{!}{%
\begin{tabular}{lcccccc}
\toprule
\multirow{2}{*}{\textbf{Mathlib Version}}
 & \multicolumn{3}{c}{\textbf{PutnamBench}}
 & \multicolumn{3}{c}{\textbf{MiniF2F}} \\
\cmidrule(lr){2-4}\cmidrule(lr){5-7}
 & LeanRefactor & ProofOptimizer & Original
 & LeanRefactor & ProofOptimizer & Original \\
\midrule
v4.20.1 & 71 & 70 & 71 & 194 & 194 & 194 \\
v4.21.0 & 69 & 70 & \textbf{71} & 193 & 194 & 194 \\
v4.22.0 & 64 & 63 & 64 & \textbf{189} & 188 & 187 \\
v4.23.0 & 51 & 50 & \textbf{61} & \textbf{175} & 170 & 151 \\
v4.24.0 & 50 & 50 & \textbf{60} & \textbf{172} & 167 & 148 \\
v4.25.0 & 49 & 48 & \textbf{58} & \textbf{168} & 162 & 141 \\
v4.26.0 & 48 & 48 & \textbf{58} & \textbf{168} & 162 & 141 \\
v4.27.0 & 45 & 45 & \textbf{53} & \textbf{167} & 162 & 140 \\
v4.28.0 & 45 & 45 & \textbf{53} & \textbf{166} & 161 & 140 \\
v4.29.0 & 43 & 42 & \textbf{52} & \textbf{166} & 161 & 140 \\
\bottomrule
\end{tabular}%
}
\vspace{-1em}
\end{table}
The comparison against original long proofs is benchmark-dependent: on MiniF2F, LeanRefactor transfers furthest among all baselines, whereas on PutnamBench it falls short of the originals and tracks ProofOptimizer closely. We therefore conclude that whether shortening substitutes tactics that transfer better or worse across releases depends on corpus composition.

\section{Proof Dependencies in Research Repositories and Competition Problems}
\label{sec:proof_dependencies_count}
\begin{table}[H]
\captionsetup{font=small}
\caption{Competition-type Lean sources rely on much fewer dependencies (Mathlib and intra-project) than Research-type Lean sources.}
\centering
\resizebox{0.6\textwidth}{!}{
\begin{tabular}{ccc}
\toprule
Type & Lean Source & Average Number of Dependencies Used \\ \midrule
\multirow{2}{*}{Competition} & miniF2F~\cite{minif2f} & 2.66 \\
 & PutnamBench~\cite{putnam} & 10.74 \\ \midrule
\multirow{5}{*}{Research} & FLT~\cite{FLT_Lean} & 39.07 \\
 & PFR~\cite{pfr_formalization} & 40.64 \\
 & PhysLean~\cite{physlib_lean} & 38.53 \\
 & Analysis~\cite{tao_lean_analysis} & 35.38 \\
 % & PDE~\cite{LeanPDERepo} & \\
\bottomrule
\end{tabular}
\label{table:theorem_used}
}
\end{table}

\section{Experimental Settings}
\label{sec:experimental_settings}
 
This section provides full details for the experiments summarized in
Section~\ref{sec:experiment}. We describe (i) the seven evaluation benchmarks and how their
initial long proofs are obtained, (ii) the metrics and the rationale for the benchmarks each
metric is computed on, (iii) the baseline configurations used for the controlled ablation and
the Claude Code comparison, and (iv) the implementation and hyperparameter settings shared
across all runs.
 
\subsection{Benchmarks}
\label{app:exp-benchmarks}
 
\paragraph{Competition problems.} We evaluate on three competition benchmarks.
 
\begin{figure*}[h!]
    \centering
    % First plot
    \begin{subfigure}{0.32\textwidth}
        \centering
        \includegraphics[width=\linewidth]{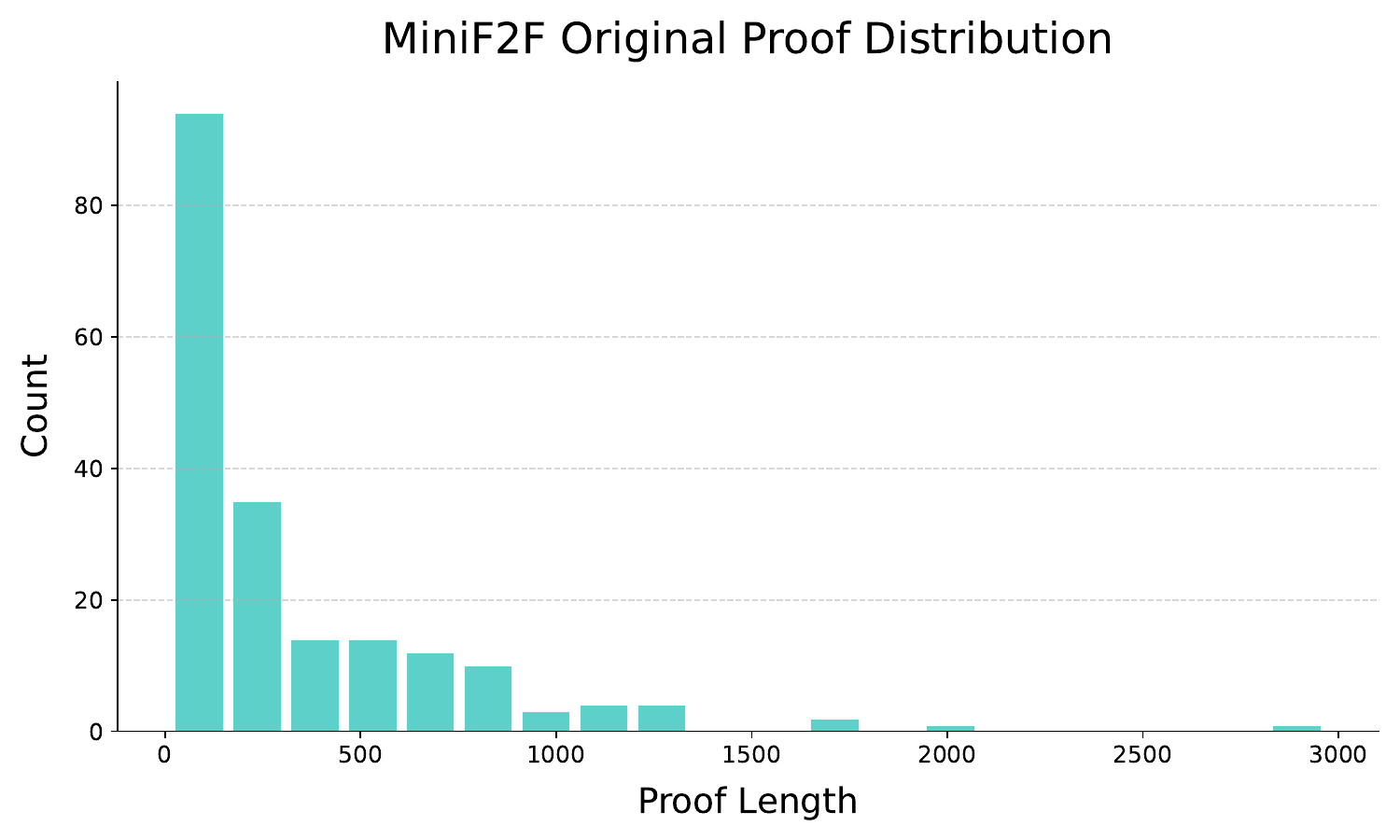}
        % \caption{miniF2F Length Distribution}
        \label{fig:comp1}
    \end{subfigure}\hfill
    % Second plot
    \begin{subfigure}{0.32\textwidth}
        \centering
        \includegraphics[width=\linewidth]{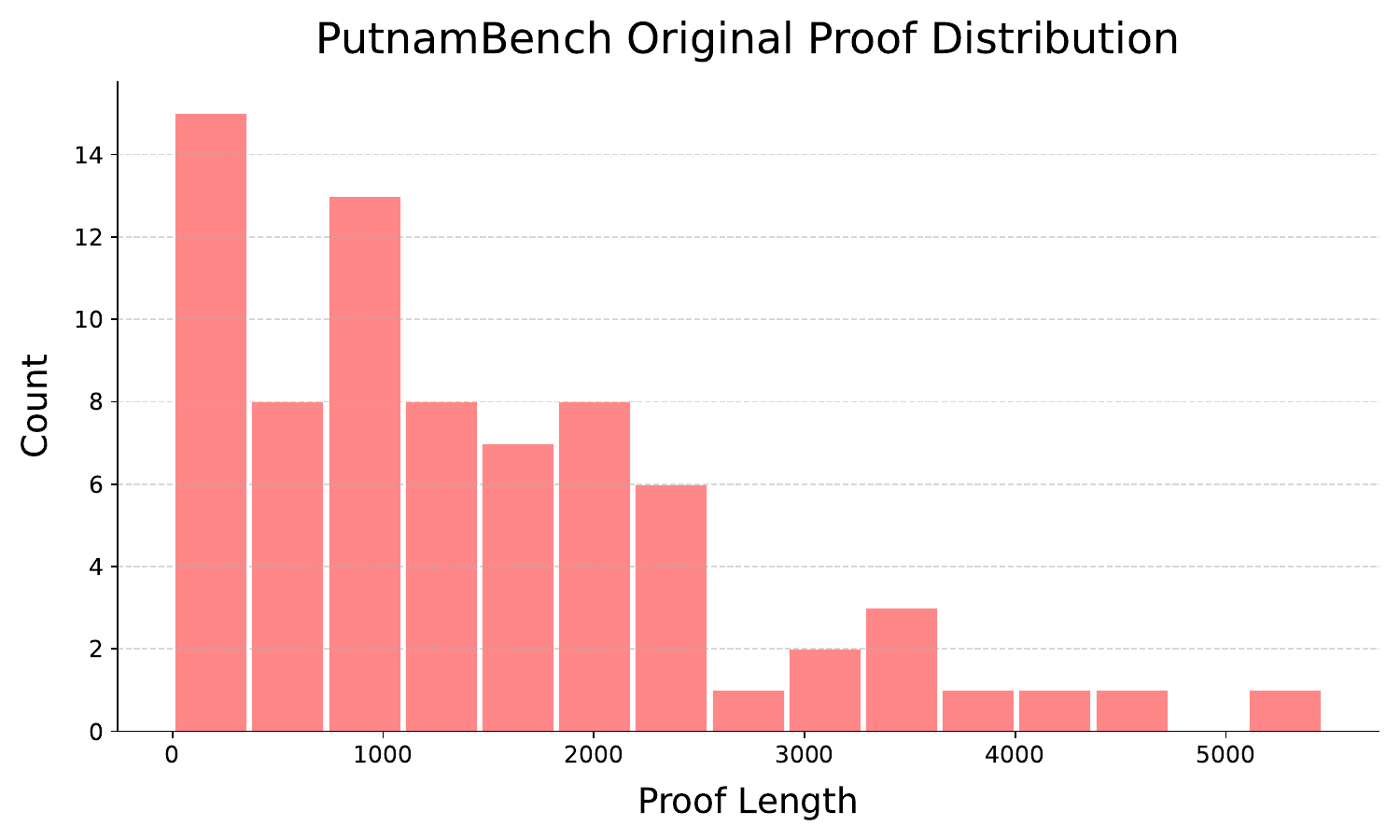}
        % \caption{Competition B}
        \label{fig:comp2}
    \end{subfigure}\hfill
    % Third plot
    \begin{subfigure}{0.32\textwidth}
        \centering
        \includegraphics[width=\linewidth]{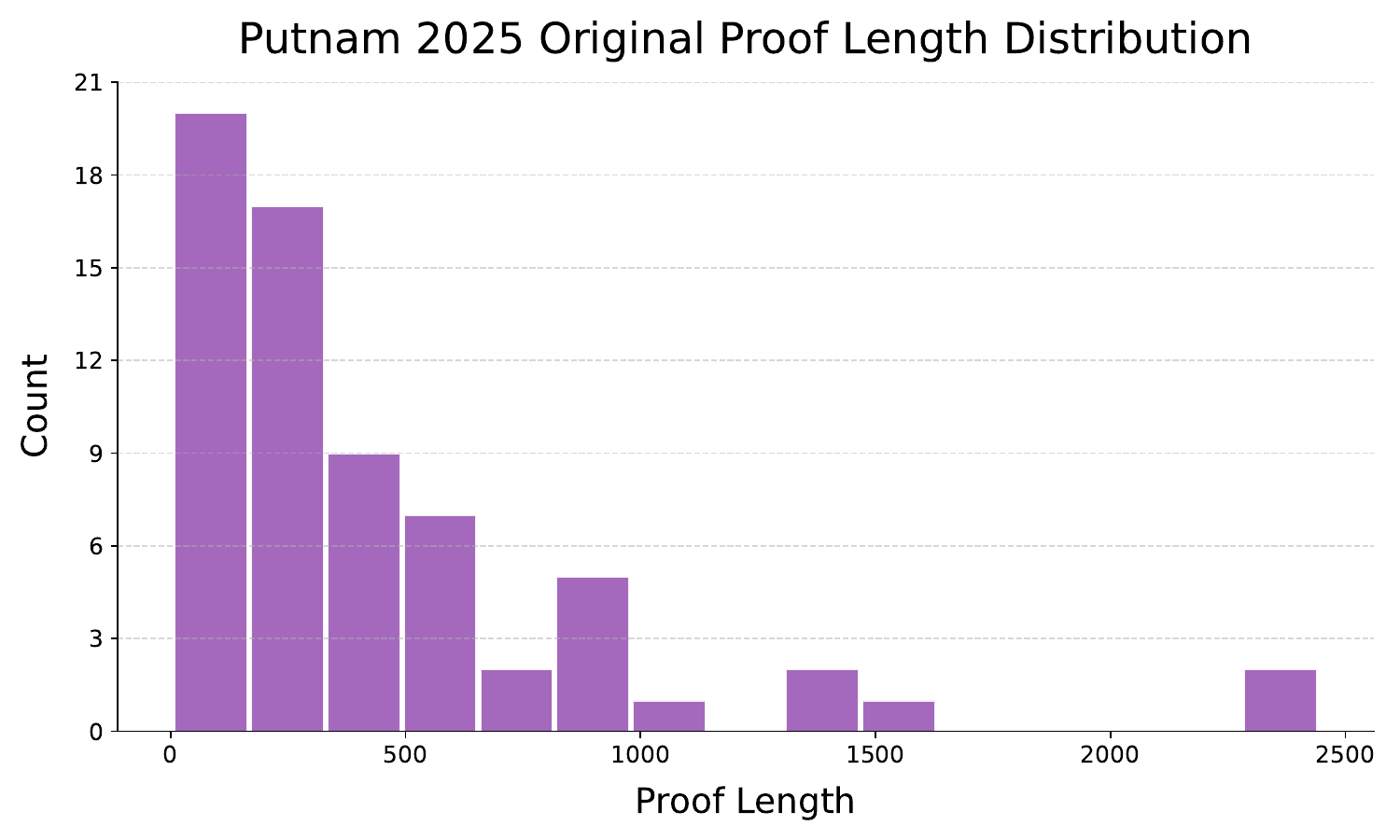}
        % \caption{Competition C}
        \label{fig:comp3}
    \end{subfigure}
    \vspace{-1em}
    
    \caption{Distribution of original proof lengths across three competition style evaluation sets (miniF2F, PutnamBench, Putnam 2025).}
    \label{fig:competition_original_distributions}
\end{figure*}

\begin{itemize}[leftmargin=*]
    \item \textbf{miniF2F}~\cite{minif2f} (194 theorems). Initial long proofs are taken
    directly from the public release of our primary baseline
    ProofOptimizer~\cite{gu2025proofoptimizer}, ensuring identical inputs across both
    methods.
    \item \textbf{PutnamBench}~\cite{putnam} (75 theorems). Initial long proofs are likewise
    taken from the ProofOptimizer release for direct comparability.
    \item \textbf{Putnam2025} (66 theorems). We construct this benchmark from the official
    Lean~4 solutions to problems B1 and B2 of the 2025 Putnam Competition produced by
    Seed-Prover~1.5~\cite{chen2025seedprover15masteringundergraduatelevel}. We restrict
    attention to B1/B2 due to budget. Each official solution is provided as a
    self-contained Lean~4 file; we extract the individual theorems and track the dependencies invoked in the original proof, yielding 66
    theorems.
\end{itemize}
The length distribution of the initial proofs across all three competition benchmarks is shown
in Figure~\ref{fig:competition_original_distributions}.

\begin{figure*}[h!]
    \centering
    % Top Left
    \begin{subfigure}{0.48\textwidth}
        \centering
        \includegraphics[width=\linewidth]{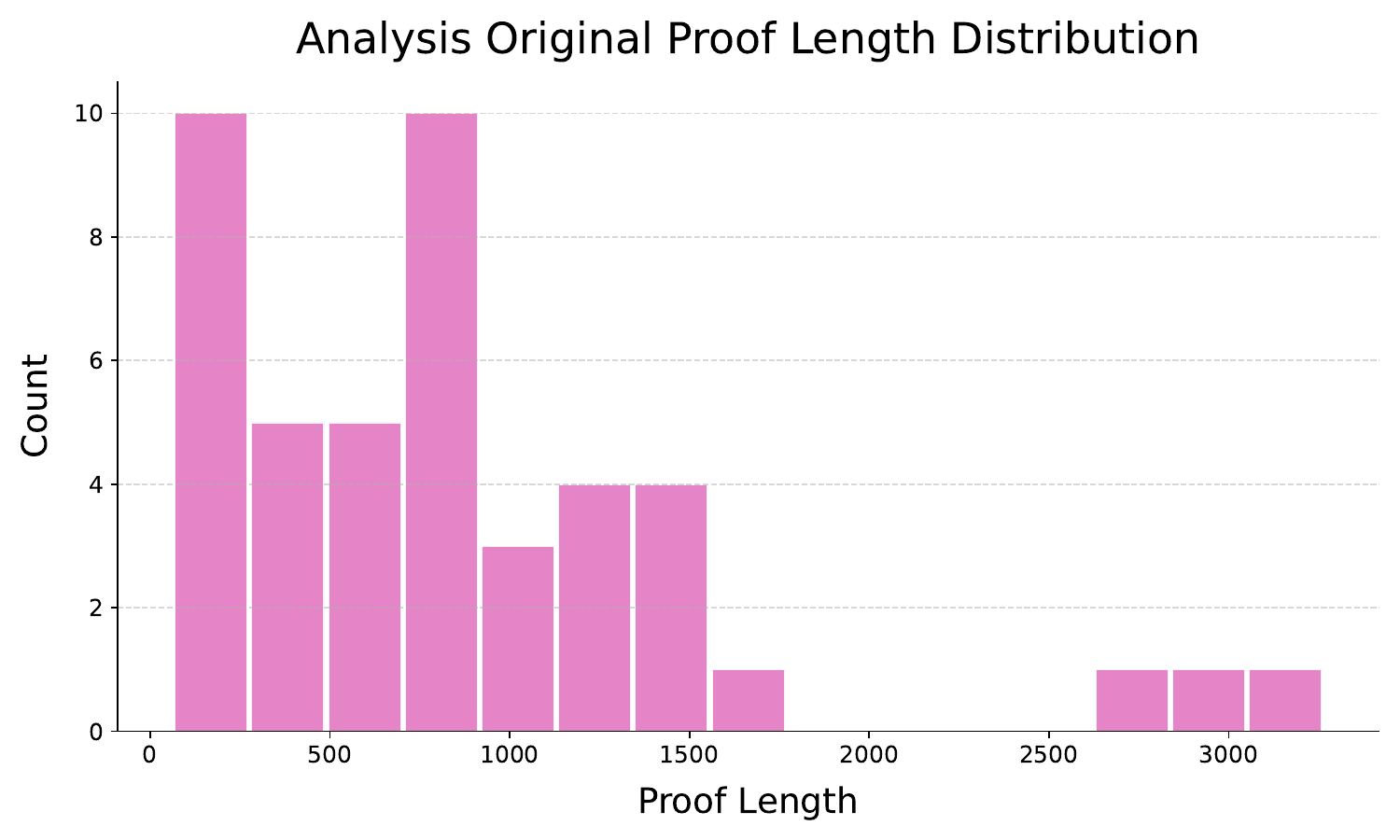}
        \label{fig:res1}
    \end{subfigure}\hfill
    % Top Right
    \begin{subfigure}{0.48\textwidth}
        \centering
        \includegraphics[width=\linewidth]{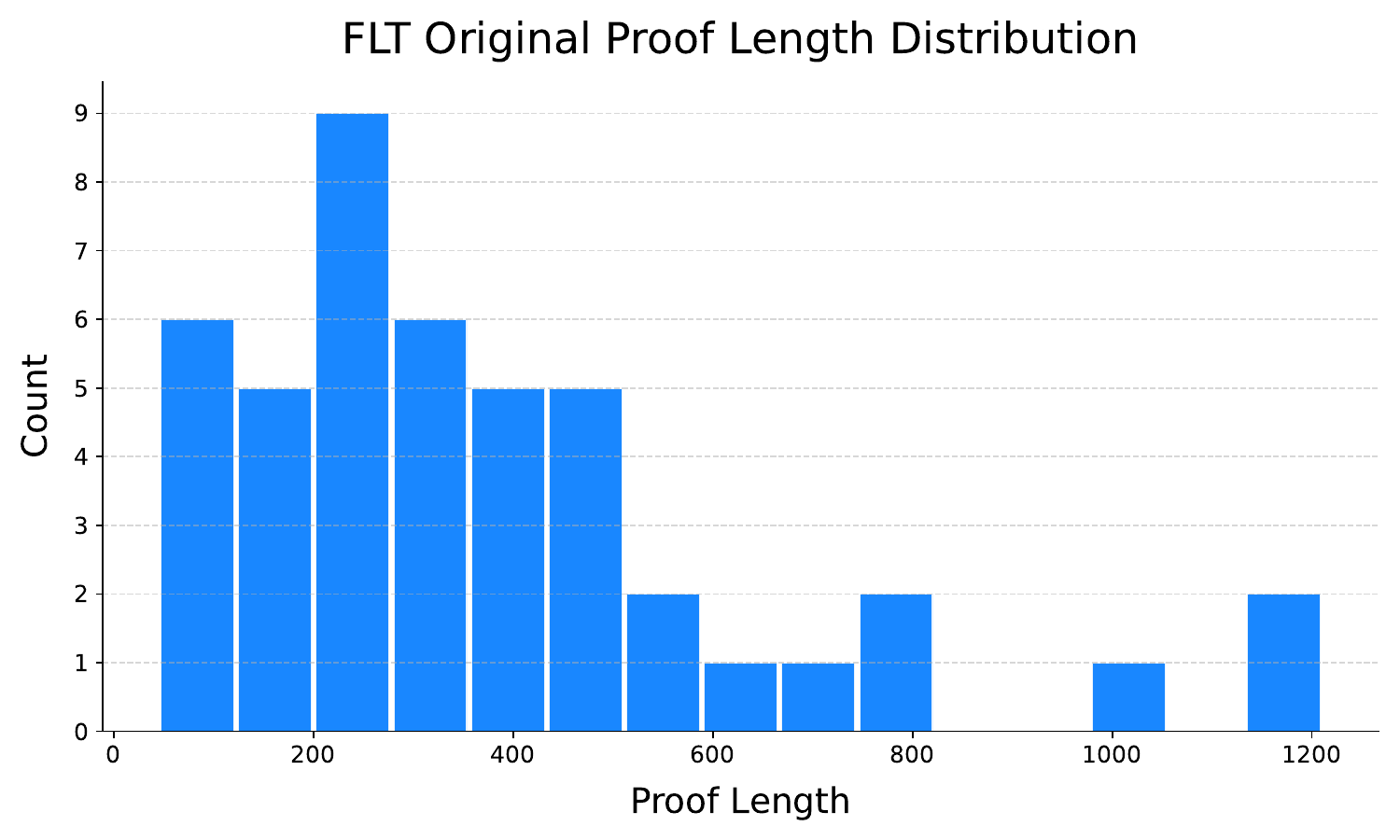}
        \label{fig:res2}
    \end{subfigure}
    
    \vspace{-1em} % Adds a nice little gap between the top and bottom rows
    
    % Bottom Left
    \begin{subfigure}{0.48\textwidth}
        \centering
        \includegraphics[width=\linewidth]{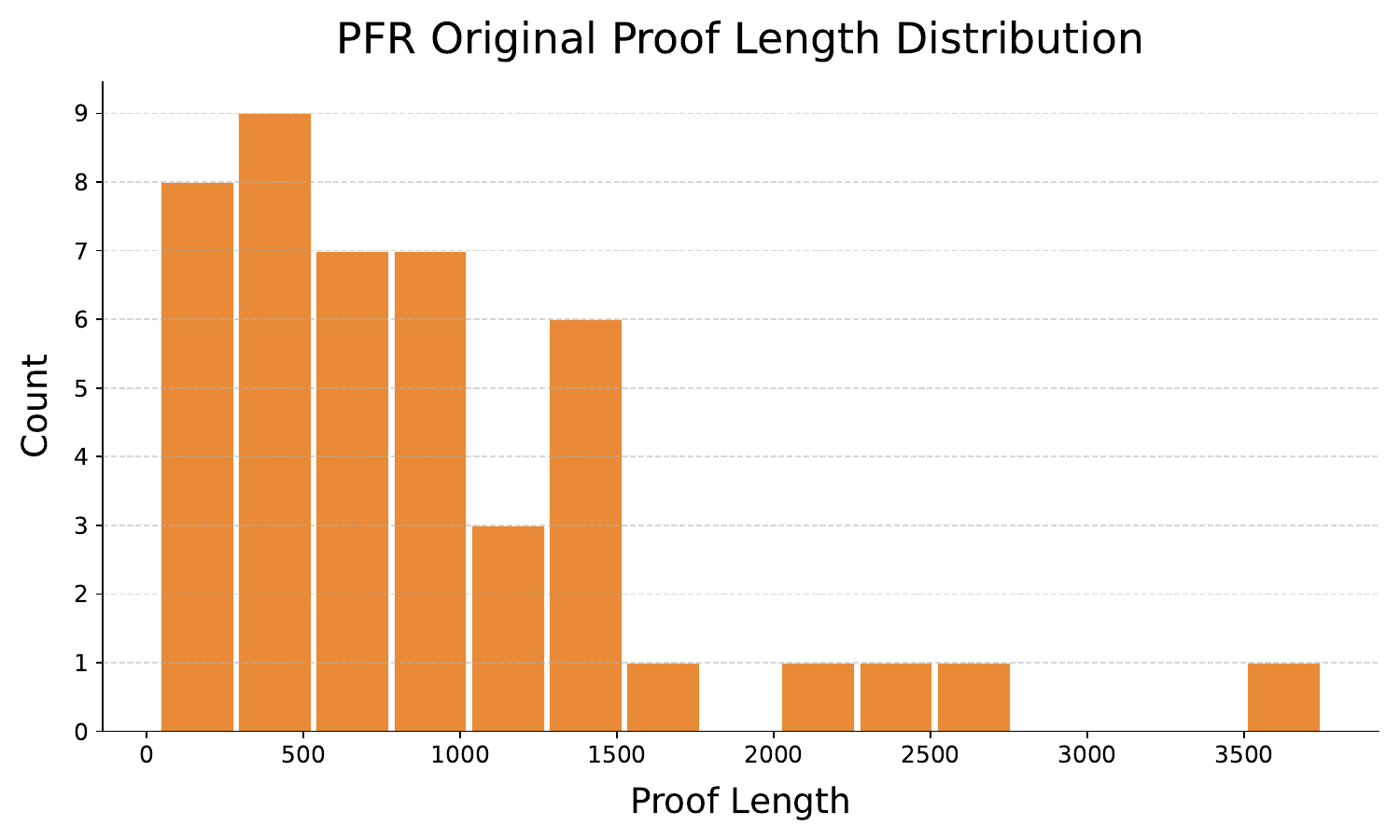}
        \label{fig:res3}
    \end{subfigure}\hfill
    % Bottom Right
    \begin{subfigure}{0.48\textwidth}
        \centering
        \includegraphics[width=\linewidth]{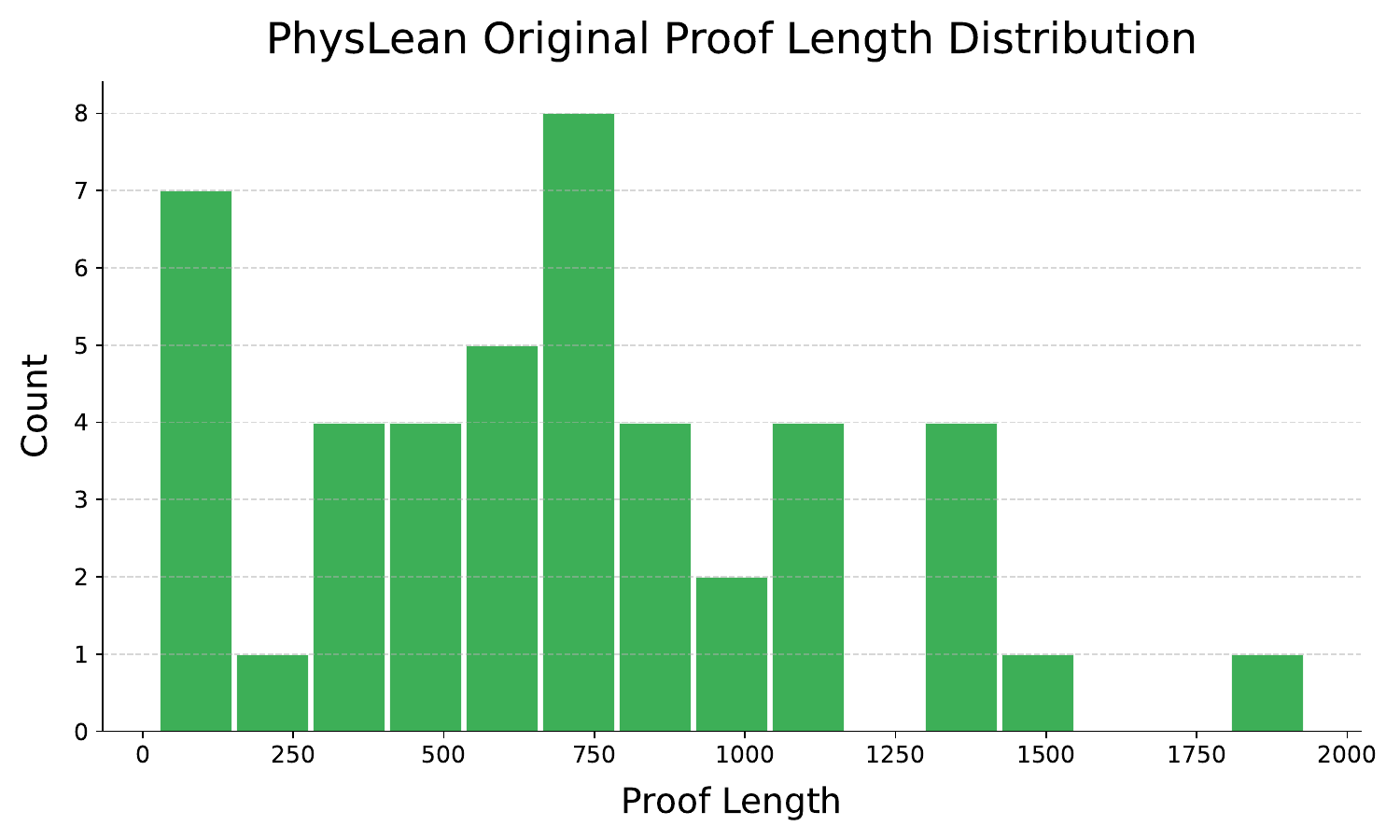}
        \label{fig:res4}
    \end{subfigure}

    \vspace{-1em}
    \caption{Distribution of original proof lengths across four research-level evaluation sets (Analysis, FLT, PFR, PhysLean).}
    \label{fig:research_original_distributions}
\end{figure*}

\paragraph{Research-level problems.} To assess refactoring on advanced mathematics, we draw
theorems from four research-level Lean~4 repositories: Analysis~\cite{tao_lean_analysis}, Fermat's Last Theorem (FLT)~\cite{FLT_Lean}, the Polynomial
Freiman--Ruzsa Conjecture (PFR)~\cite{pfr_formalization}, and PhysLean~\cite{physlib_lean}.
From each repository we sample 45 theorems, subject to compute budget. To evaluate refactoring
effectiveness across a wide range of initial proof complexities, we stratify the sample by
proof length:
\begin{itemize}[leftmargin=*]
    \item 15 proofs exceeding 1000 tokens,
    \item 15 proofs between 500 and 1000 tokens,
    \item 10 proofs between 100 and 500 tokens,
    \item 5 proofs under 100 tokens.
\end{itemize}
When a repository contains insufficient theorems in a given length bucket, the unfilled quota
is reallocated to the nearest non-empty bucket. The extraction pipeline retains all internal
and external dependencies, ensuring that the agent operates on a properly contextualized,
dependency-aware proof environment. The length distribution of the initial proofs across the
four research-level evaluation sets is shown in
Figure~\ref{fig:research_original_distributions}.

\subsection{Metrics}
\label{app:exp-metrics}
 
We report three metrics across our benchmarks.
 
\paragraph{Proof length.} We measure proof brevity as the average relative percentage decrease
in token count between the initial long proof and the refactored proof. To ensure consistency
with prior work, we use the syntax-aware tokenizer released by
ProofOptimizer~\cite{gu2025proofoptimizer}; code is provided in
Appendix~\ref{sec:proof_length_code}. This metric is reported on all seven benchmarks.

\paragraph{Compilation cost.} We assess computational efficiency using \texttt{lake env lean --profile}, which exposes per-declaration timing. We isolate proof execution time from library import overhead, since import time is dominated by caching effects unrelated to proof content. Each measurement is taken in a consistent environment to ensure stable, deterministic timings, and we report the mean over five runs on identical hardware. On miniF2F and PutnamBench we follow the per-proof setting as each proof is a standalone \texttt{.lean} file, compiling each refactored proof in isolation. Research repositories do not admit clean per-proof wall-clock attribution, so we instead measure heartbeats, Lean 4's deterministic proxy for elaboration work; see Appendix~\ref{sec:compilation_time_research_level} for full details. This metric is reported on all seven benchmarks.
 
\paragraph{Cross-version compatibility.} To evaluate the impact of our version-tagged
metadata, we compare the agent's performance under \emph{version-filtered} strategy retrieval
against an unfiltered baseline. Refactoring is run across three Lean and \texttt{Mathlib}
toolchains (\texttt{v4.14.0}, \texttt{v4.16.0}, and \texttt{v4.22.0}); we then measure proof
length reduction in each environment to determine whether retrieval tailored to a target
version yields larger gains than version-agnostic retrieval. We additionally assess zero-shot
version transfer by recompiling \texttt{v4.19.0} PutnamBench outputs from Lean Refactor, ProofOptimizer,
and the original verbose proofs against later \texttt{Mathlib} releases through
\texttt{v4.29.0} without further refactoring, measuring how many proofs continue to
type-check across future toolchains.

\subsection{Baselines}
\label{app:exp-baselines}
 
\paragraph{ProofOptimizer.} On miniF2F and PutnamBench, we benchmark against
ProofOptimizer~\cite{gu2025proofoptimizer} using their officially released shortest proofs,
obtained via an 8-iteration shortening process with sampling budget $64$ in the first six
iterations and $1024$ in the final two (denoted $6{\times}64+2{\times}1024$). Because the
ProofOptimizer model checkpoints are closed-source, we cannot run their model on the
remaining benchmarks (Putnam2025, Analysis, FLT, PFR, PhysLean). On all benchmarks, we conduct a controlled ablation over variants of our own framework.
 
\paragraph{Internal ablations.} We define four internal baselines, each ablating a different
component of Lean Refactor:
 
\begin{itemize}[leftmargin=*]
    \item \textbf{Base Agent.} The framework stripped to its core execution loop. The Planner
    module is removed; only the Refactorer and the verifier-guided Debugger remain. Strategy
    retrieval is applied as in our full method (Section~\ref{sec:agent_description}), and the
    retrieved strategies are passed directly to the Refactorer. \textit{This isolates the
    contribution of the Planner.}
    \item \textbf{Base Agent w/o Retrieval.} The Base Agent, with strategy retrieval (Section~\ref{sec:multi_objective_retrieval}) additionally disabled. \textit{This isolates the contribution of
    retrieval within the simplified loop.}
    \item \textbf{Lean Refactor w/o Retrieval.} The complete agent architecture (Planner,
    Refactorer, and Debugger) with the retrieval module disabled. \textit{This isolates the gains
    attributable purely to the Planner's task decomposition, in the absence of external
    strategy context.}
    \item \textbf{Lean Refactor w/ Random Retrieval.} The full agent framework, but with our
    semantic and metadata-aware retrieval replaced by uniform random sampling from the
    strategy bank. This serves as a \textit{control for the embedding-and-rerank pipeline itself},
    distinguishing the value of \emph{retrieving the right} strategies from the value of
    \emph{having any} strategies in context.
\end{itemize}
 
\paragraph{LLM-agnosticism and Claude Code comparison.} We additionally run our framework
with Claude Haiku 4.5~\cite{anthropic2025claude45haiku} to test that gains are not specific
to Gemini~3~Flash. Using the same Claude Haiku 4.5 backbone, we compare against
\textbf{Claude Code}. We
prompt Claude Code with the same proof-shortening objective as our Lean Refactor agent,
but adapted to its native abstractions (skills and MCP servers) rather than forcing it into
our planner--refactorer--debugger loop. This lets Claude Code exercise its own agentic
control flow on the task our framework is designed for, isolating the contribution of our
architecture from the contribution of simply prompting a capable general-purpose agent to shorten Lean
proofs. We evaluate Claude Code under three tool configurations of increasing support:
\begin{itemize}[leftmargin=*]
    \item \textbf{Lean MCP only.} The agent is provided only with the Lean MCP
    server\footnote{\url{https://github.com/oOo0oOo/lean-lsp-mcp}}, which exposes compiler
    feedback for correctness checking but no refactoring guidance.
    \item \textbf{Lean MCP + Lean skills.} The previous configuration is augmented with the
    Lean skills package\footnote{\url{https://github.com/cameronfreer/lean4-skills}}, which
    supplies general Lean~4 refactoring and golfing guidance.
    \item \textbf{Lean MCP + Lean skills + our strategy retrieval tool.} The previous configuration is
    further augmented with a retrieval tool that exposes our strategy bank as an on-demand
    resource the agent can query during refactoring.
\end{itemize}
This comparison isolates the marginal value contributed by our strategy bank when added to a
strong off-the-shelf coding agent. Due to compute budget, we run the Claude Code comparison
on one representative benchmark per regime: PutnamBench (competition) and Analysis
(research-level).
 
\subsection{Implementation Details}
\label{app:exp-implementation}
 
\paragraph{Lean environment.} All data collection and all experiments use Lean
\texttt{v4.24.0} as the default environment, except where the cross-version compatibility
study explicitly requires \texttt{v4.14.0}, \texttt{v4.16.0}, or \texttt{v4.22.0}.
 
\paragraph{Model serving.} All open-weight models used during data collection (in particular GPT-OSS and the Qwen3-Embedding-8B) are primarily served via vLLM with default sampling parameters. Additionally, we utilized the OpenRouter API for GPT-OSS. For all GPT-OSS deployments, we configured the model using the medium reasoning-strength setting. Total compute for the reported experiments was approximately 10{,}000 GPU-hours on NVIDIA 6000 Ada GPUs, with strategy-bank construction (proof synthesis, strategy distillation, deduplication) accounting for the majority of this cost. API-served models (GPT-OSS-120B via OpenRouter, Gemini 3 Flash, and Claude Haiku 4.5) incurred an additional cost of approximately \$5{,}000 USD. 
 
\paragraph{Agent configuration.} For our agent framework, we use:
\begin{itemize}[leftmargin=*]
    \item a maximum of 3 debugging cycles per refactoring episode ($D=3$),
    \item a total API call budget of 30 per theorem ($B=30$),
    \item the minimum proof length during optimization is 5 ($T=5$),
    \item proof chunking with chunk sizes of 5, 10, and 20 lines. The multi-granularity
    chunking ensures that retrieval queries target proof regions of different lengths and
    consequently surface strategies appropriate to refactoring at different granularities.
    \item experiments were run on servers equipped with Intel Xeon w9-3495X CPUs and NVIDIA 6000 Ada GPUs.
\end{itemize}
 
\paragraph{Backbone LLMs.} Unless stated otherwise, all main-paper experiments use
Gemini~3~Flash~\cite{geminiteam2025gemini3flash} as the agent backbone. Claude Haiku
4.5~\cite{anthropic2025claude45haiku} is used both for the LLM-agnosticism check and as the
shared backbone in the Claude Code comparison.

\section{Code for Getting Proof Length}
\label{sec:proof_length_code}
\begin{lstlisting}[style=mypython]
def proof_length(statement_and_proof):
    """
    Compute the token count of a proof from a full statement string.
    Extracts the proof by finding where the signature ends, then tokenizes and counts.
    """
    lean_operators = [
        ":=", "!=", "&&", "-.", "->", "←", "..", "...", "::", ":>", 
        "<;>", ";;", "==", "||", "=>", "<=", ">=", "⁻¹", "?_"
    ]
    lean_operators_spaced = [" ".join(conn) for conn in lean_operators]
    lean_operators_dict = dict(zip(lean_operators_spaced, lean_operators, strict=False))

    def lexer(lean_snippet):
        tokenized_lines = []
        for line in lean_snippet.splitlines():
            tokens = []
            token = ""
            for ch in line:
                if ch == " ":
                    if token:
                        tokens.append(token)
                        token = ""
                elif str.isalnum(ch) or (ch in "._'"):
                    token += ch
                else:
                    if token:
                        tokens.append(token)
                        token = ""
                    tokens.append(ch)
            if token:
                tokens.append(token)
            tokenized_line = " ".join(tokens)
            for conn in lean_operators_spaced:
                if conn in tokenized_line:
                    tokenized_line = tokenized_line.replace(conn, lean_operators_dict[conn])
            tokenized_lines.append(tokenized_line)
        return "\n".join(tokenized_lines)

    try:
        statement_and_proof = remove_comments(statement_and_proof) # Remove all comments in the proof
        decl_start, decl_end = return_theorem_to_prove(statement_and_proof) # Gets the declaration start and end
        proof = statement_and_proof[decl_end:] # Isolate the proof from the declaration
        proof_tokenized = lexer(proof)
        return sum([len(l.split(" ")) for l in proof_tokenized.splitlines()])
    except:
        return 10**9
\end{lstlisting}

\section{Comparison with ProofOptimizer's Heartbeat-Optimized Proofs}
\label{sec:heartbeat_comparison}

The wall-clock results in Table~\ref{tab:compilation_time_combined} compare against ProofOptimizer's
\emph{length-optimized} proofs because their heartbeat-optimized model weights are not
released, putting their best-case compilation numbers out of reach of direct re-execution.
We close this gap using Lean's \texttt{\#count\_heartbeats} (under
\texttt{set\_option Elab.async false}) as a deterministic proxy of
compilation cost, the same metric ProofOptimizer adopts when optimizing for
execution efficiency. This permits a direct comparison against the compilation cost numbers that ProofOptimizer~\cite{gu2025proofoptimizer} publish: the values produced by their heartbeat-optimized inference-time variant after eight iterations with sampling budget $6{\times}64 + 2{\times}1024$, i.e.\ their strongest reported heartbeat results. We
compute the heartbeats for proofs refactored under the \emph{Lean Refactor w/ Reranking} setting from Section~\ref{sec:results_compilation_time}, and report the comparison in Table~\ref{tab:heartbeat_comparison}.

\begin{table}[h]
\centering
\caption{\textbf{Heartbeat-level comparison against ProofOptimizer's heartbeat-optimized
variant.} Heartbeats reports the average heartbeat count (in thousands) across theorems. ProofOptimizer values are taken from their heartbeat-optimized proofs; ``Original'' refers to the unrefactored proofs from the same source. Lower is better; best per column in bold.}
\label{tab:heartbeat_comparison}
\resizebox{\textwidth}{!}{%
\begin{tabular}{lcc}
\toprule
\textbf{Method} & \textbf{miniF2F Heartbeats (K)} $\downarrow$ & \textbf{PutnamBench Heartbeats (K)} $\downarrow$ \\
\midrule
Original (no refactoring) & 36.3 & 221 \\
ProofOptimizer, Heartbeat-optimized~\cite{gu2025proofoptimizer} & 10.4 & 111 \\
\textbf{Lean Refactor w/ Reranking} (Ours) & \textbf{9.2} & \textbf{82} \\
\bottomrule
\end{tabular}%
}
\end{table}

\paragraph{Objective-conditioned reranking yields the lowest heartbeat counts.}
Across both benchmarks, Lean Refactor w/ Reranking attains the lowest average heartbeat
count among all evaluated methods, improving over ProofOptimizer's heartbeat-optimized
proofs from $10.4$K to $9.2$K on miniF2F and from $111$K to $82$K on PutnamBench. We
emphasize that ProofOptimizer's underlying model is trained to generate shorter proofs;
its heartbeat-optimized variant achieves the published numbers by reusing the same
length-trained model and selecting low-heartbeat candidates at inference. This exposes a
train--inference objective mismatch: the prior is shaped by token length, while the
selection rule targets execution cost. Our framework avoids this mismatch by carrying the
objective in the retrieval rule rather than in model weights, reranking a shared,
semantically-retrieved candidate pool by per-strategy heartbeat metadata
(Section~\ref{sec:multi_objective_retrieval}). Because the metadata is annotated directly on
strategies, the same bank can be redirected to new cost metrics without retraining.

\section{Compilation Cost for Research-Level Proofs}
\label{sec:compilation_time_research_level}
 
For competition benchmarks, each proof lives in a self-contained \texttt{.lean} file, so we
can isolate proof execution from import overhead and report wall-clock time directly. Research
repositories invert this setup: proofs reside inside large multi-file Lean projects whose
elaboration is interleaved with import resolution, neighboring declarations, and kernel cache
state, and there is no stable way to attribute wall-clock
time to a single proof in this regime\footnote{\url{https://github.com/leanprover/lean4/issues/8038}}. We
therefore switch to \emph{heartbeats}, Lean 4's deterministic proxy for elaboration work. We prepend each proof
with \texttt{set\_option Elab.async false in} and \texttt{\#count\_heartbeats in} to pin the
count to the declaration under measurement, consistent with prior work~\cite{gu2025proofoptimizer}. We report dataset-level means of the per-proof absolute and relative reductions in Table~\ref{tab:heartbeats_research}.

\begin{table}[h]
\centering
\caption{\textbf{Heartbeat reduction on Analysis~\cite{tao_lean_analysis}, FLT~\cite{FLT_Lean},
pfr~\cite{pfr_formalization}, and PhysLean~\cite{physlib_lean}.}
\emph{Avg. Rel. Reduc.}: average per-proof relative \% decrease in heartbeats.
\emph{Avg. Abs. Reduc.}: average per-proof absolute decrease in heartbeats. Higher is better
for both. Each dataset contains $n=45$ proofs.}
\label{tab:heartbeats_research}
\resizebox{\linewidth}{!}{%
\begin{tabular}{llcc}
\toprule
\textbf{Dataset} & \textbf{Method} & \textbf{Avg. Rel. Reduc. $\uparrow$ (\%)} & \textbf{Avg. Abs. Reduc. $\uparrow$ (heartbeats)} \\
\midrule
\multirow{4}{*}{Analysis~\cite{tao_lean_analysis}}
 & Lean Refactor w/o Retrieval         & $-6.72$          & $2324$ \\
 & Lean Refactor w/ Random Retrieval   & $-4.18$          & $2759$ \\
 & Lean Refactor                       & $-14.43$         & $3049$ \\
 & \textbf{Lean Refactor w/ Reranking}          & $\mathbf{3.19}$  & $\mathbf{3809}$ \\
\midrule
\multirow{4}{*}{FLT~\cite{FLT_Lean}}
 & Lean Refactor w/o Retrieval         & $0.37$           & $-946$ \\
 & Lean Refactor w/ Random Retrieval   & $0.70$           & $222$ \\
 & Lean Refactor                       & $3.00$           & $311$ \\
 & \textbf{Lean Refactor w/ Reranking}          & $\mathbf{11.33}$ & $\mathbf{1235}$ \\
\midrule
\multirow{4}{*}{PFR~\cite{pfr_formalization}}
 & Lean Refactor w/o Retrieval         & $-6.81$          & $1332$ \\
 & Lean Refactor w/ Random Retrieval   & $9.21$           & $3223$ \\
 & Lean Refactor                       & $3.13$           & $2184$ \\
 & \textbf{Lean Refactor w/ Reranking}          & $\mathbf{9.89}$  & $\mathbf{3309}$ \\
\midrule
\multirow{4}{*}{PhysLean~\cite{physlib_lean}}
 & Lean Refactor w/o Retrieval         & 1.41             & 1152 \\
 & Lean Refactor w/ Random Retrieval   & 3.73             & 2166 \\
 & Lean Refactor                       & -21.97             & 2609 \\
 & \textbf{Lean Refactor w/ Reranking}          & $\mathbf{7.47}$    & $\mathbf{4101}$ \\
\bottomrule
\end{tabular}
}
\vspace{-0.5em}
\end{table}
 
\textbf{Lean Refactor with reranking is the strongest method by both metrics on every dataset}, indicating that objective-aligned retrieval, not structural rewriting alone, is what consistently delivers heartbeat savings. The two columns can disagree in sign (notably for Analysis, PFR, and PhysLean under non-reranking variants) because the relative mean is sensitive to outliers: a few refactored proofs exhibit large heartbeat increases that pull the average relative reduction negative, even when the absolute mean remains positive and the majority of proofs improve.

\section{Additional Proof Length Reduction Results}
\label{sec:extra_length_reduction_results}

We further evaluate Lean Refactor on two additional benchmarks from distinct domains: \textbf{Verina}~\cite{ye2026verinabenchmarkingverifiablecode}, a benchmark for verifiable code generation in Lean, from which we use the proof-generation task, and a \textbf{PDE}~\cite{Stehling2026Lean} benchmark of formalized statements about partial differential equations.

\paragraph{Verina setup.} Verina consists of 189 problems. Since no public set of model-generated proofs is available, we first obtain solutions with a refinement-loop theorem prover built on top of Claude Opus 4.6: for each task the model generates a candidate proof, the Lean compiler checks it, and compiler errors are fed back as additional context, with up to 16 attempts at temperature 1.0 (matching the Verina default). Each candidate is validated with integrity checks (no modification to the original code, specification, or theorem statement) and axiom checks (only \texttt{propext}, \texttt{Classical.choice}, and \texttt{Quot.sound} are permitted). This solver closes 14/189 problems (7.4\%; 1 advanced and 13 basic tasks), yielding 15 theorems to which we then apply Lean Refactor.

\paragraph{PDE setup.} We follow the same sampling strategy specified in Appendix~\ref{app:exp-benchmarks} and sample 45 theorems spanning diverse proof lengths.

\paragraph{Results.} Lean Refactor framework achieves an average relative proof-length reduction of \textbf{53.50\%} on Verina and \textbf{35.34\%} on PDE. This indicates that the framework transfers cleanly from competition mathematics to verifiable code generation (Verina~\cite{ye2026verinabenchmarkingverifiablecode}) and to formalized PDE topics (PDE~\cite{Stehling2026Lean}), supporting the claim that Lean Refactor's refactoring performance reflects a general capability for optimizing Lean proofs across mathematical and verification domains.

\section{Examples of Refactoring Strategies}
\label{sec:strategy_examples}

\subsection*{Strategy 1: Eliminate exhaustive numeric case analysis; construct new witness via arithmetic shift}
\noindent\textbf{Description.} A wasteful proof pattern: extract
concrete numeric values for an existential witness (and any auxiliary
constants) by evaluating a universal hypothesis at many points and
discharging the resulting system with \texttt{norm\_num},
\texttt{nlinarith}, and \texttt{linarith}. This is unnecessary when
the goal differs from the hypothesis only by a shift in the bound
variable --- the witness can be obtained by translating the existing
one algebraically, leaving the numeric values abstract.
\medskip

\noindent\textbf{When to apply.} The hypothesis has the form
\texttt{$\exists$ x, $\forall$ y, P x y} and the goal has the form
\texttt{$\exists$ x, $\forall$ y, P x (y - k)} (or any other affine
reparametrisation of \texttt{y}). \textbf{Note:} The remaining goal
after providing the shifted witness must close by specializing the
hypothesis at the shifted point --- i.e.\ it must not depend on the
specific numeric value of the witness. If it does, the original
numeric derivation is needed.
\medskip

\noindent\textbf{Application guide.}
\begin{enumerate}
  \item Delete every block that pins the witness or auxiliary
        constants to specific numeric values --- the long chains of
        \texttt{have h\_i := hyp v\_i} followed by \texttt{norm\_num},
        \texttt{nlinarith}, \texttt{linarith}.
  \item Destructure the existential: \texttt{rcases h with $\langle$a,
        ha$\rangle$}.
  \item Provide the shifted witness:
        \texttt{refine $\langle$a + k, ?\_$\rangle$}, where $k$ is the
        shift between the hypothesis's bound variable and the goal's
        (read off the goal: if it mentions \texttt{y - k}, use
        \texttt{a + k}).
  \item Close the remaining goal by specializing \texttt{ha} at the
        shifted point, typically
        \texttt{intro y; simpa using ha (y - k)}, or a short
        \texttt{linarith}/\texttt{ring} after substitution.
\end{enumerate}
\medskip
\noindent\textbf{Example.} Schematic shape: from a hypothesis
\texttt{h : $\exists$ x, $\forall$ y, P x y} prove the shifted goal
\texttt{$\exists$ x, $\forall$ y, P x (y - k)}. The before-proof
wastefully derives the concrete value of the witness $a$ (and a
secondary constant $c$) by instantiating \texttt{h} at many integers.

\noindent\textit{Before}
\begin{lstlisting}
-- pin the witness to a concrete value
have h_a : a = a0 := by
  have e1 := h v1
  have e2 := h v2
  -- ... many more instantiations ...
  have e_n := h v_n
  norm_num at e1 e2 ... e_n
  nlinarith
-- pin an auxiliary constant
have h_c : c = c0 := by
  have e1 := h v1_prime
  -- ... many more instantiations ...
  have e_m := h v_m_prime
  norm_num [h_a] at e1 ... e_m |-
  <;> linarith
use a0 + k
-- ... finish using h_a, h_c ...
\end{lstlisting}
\noindent\textit{After}
\begin{lstlisting}
rcases h with <a, ha>
refine <a + k, ?_>
intro y
simpa using ha (y - k)
\end{lstlisting}
\medskip
\noindent\textbf{Reduction:} large.

\subsection*{Strategy 2: Replace per-entry \texttt{have} lemmas with a single \texttt{ext} and \texttt{fin\_cases} proof}
\noindent\textbf{Description.} The long proof proves each matrix entry
separately using \texttt{have} statements and then glues them together
with \texttt{Matrix.ext} inside a final \texttt{h\_main}. For matrices
indexed by a finite type, equality can be reduced to pointwise equality
by \texttt{ext}, after which a case split on the indices
(\texttt{fin\_cases}) makes each entry goal trivial. A single
\texttt{simp} (with the right lemmas) then closes all cases, removing
the need for separate per-entry lemmas.
\medskip

\noindent\textbf{When to apply.} Use this pattern whenever you are
proving equality of two matrices (or similar objects) indexed by
\texttt{Fin n} or another finite type, and you have written individual
lemmas for each entry. \textbf{Note:} Requires that the target matrix
is built via \texttt{Matrix.of} / \texttt{!![\dots]} notation (so
\texttt{Matrix.of\_apply} reduces it) and that the per-entry goals
reduce to something a unified finisher can solve. If the original
per-entry proofs required nontrivial work, that work must still be
expressible as part of the unified finisher.
\medskip

\noindent\textbf{Application guide.}
\begin{enumerate}
  \item Delete all \texttt{have} statements that compute each entry
    individually, including the final \texttt{h\_main} that wraps
    \texttt{apply Matrix.ext}.
  \item After \texttt{intro X}, write \texttt{ext i j}.
  \item Immediately follow with \texttt{fin\_cases i <;> fin\_cases j}
    to enumerate all index pairs.
  \item Finish with
\begin{verbatim}
simp [Matrix.mul_apply, Matrix.of_apply, Fin.sum_univ_two]
\end{verbatim}
    optionally followed by \texttt{<;> ring} or \texttt{<;> norm\_num}
    if arithmetic remains.
  \item Remove the final \texttt{exact h\_main} as the goal is already
    solved.
\end{enumerate}
\medskip
\noindent\textbf{Example.}
\noindent\textit{Before}
\begin{lstlisting}
intro X
have h1 : ((M : Matrix (Fin 2) (Fin 2) R) * X) 0 0 = a := by
  simp [Matrix.mul_apply]
  -- many try tactics
have h2 : ((M : Matrix (Fin 2) (Fin 2) R) * X) 0 1 = b := by
  simp [Matrix.mul_apply]
  -- many try tactics
have h3 : ((M : Matrix (Fin 2) (Fin 2) R) * X) 1 0 = c := by
  simp [Matrix.mul_apply]
  -- many try tactics
have h4 : ((M : Matrix (Fin 2) (Fin 2) R) * X) 1 1 = d := by
  simp [Matrix.mul_apply]
  -- many try tactics
have h_main : (M : Matrix (Fin 2) (Fin 2) R) * X = M2 := by
  apply Matrix.ext
  intro i j
  fin_cases i <;> fin_cases j
  simp [h1, h2, h3, h4]
exact h_main
\end{lstlisting}
\noindent\textit{After}
\begin{lstlisting}
intro X
ext i j
fin_cases i <;> fin_cases j <;>
  simp [Matrix.mul_apply, Matrix.of_apply, Fin.sum_univ_two]
\end{lstlisting}
\medskip
\noindent\textbf{Reduction:} medium.

\begin{algorithm}[tb]
\caption{Retrieval-Augmented Agent Framework for Proof Refactoring}
\label{alg:agent_framework}
\begin{algorithmic}[1]
\Require Initial unoptimized proof $P$, Strategy DB $\mathcal{S}$, Budget $B$, Target length threshold $T$, Max debug rounds $D$, Retrieval budget $K$
\Ensure Optimized refactored proof $P^*$
\State $P_{curr} \gets P$
\State $History \gets \emptyset$
\State $steps \gets 0$ 
\While{$steps < B$ \textbf{and} $\textsc{Length}(P_{curr}) > T$}
    \State $Chunks \gets \textsc{SegmentProof}(P_{curr})$
    \State $\mathcal{C}_{retrieved} \gets \emptyset$
    \For{$c \in Chunks$}
        \State $E_c \gets \textsc{Embed}(c)$
        \State $S_c \gets \textsc{RetrieveAndRerank}(\mathcal{S}, E_c, K)$ \Comment{Multi-objective alignment}
        \State $\mathcal{C}_{retrieved} \gets \mathcal{C}_{retrieved} \cup \{ (S_c, \textsc{Location}(c)) \}$
    \EndFor
    
    \State $Plan \gets \textsc{Planner}(P_{curr}, \mathcal{C}_{retrieved}, History)$
    \If{$Plan = \emptyset$}
        \State \textbf{break} \Comment{Terminate if no refactoring steps are viable}
    \EndIf
    
    \State $step\_applied \gets \textbf{False}$
    \For{$step \in Plan$}
        \State $P_{cand} \gets \textsc{Refactorer}(P_{curr}, step)$
        \State $success, \epsilon \gets \textsc{CompileLean}(P_{cand})$
        \State $d \gets 0$
        
        \Comment{Verifier-guided self-debugging loop}
        \While{\textbf{not} $success$ \textbf{and} $d < D$}
            \State $P_{cand} \gets \textsc{Debugger}(P_{cand}, \epsilon, step)$
            \State $success, \epsilon \gets \textsc{CompileLean}(P_{cand})$
            \State $d \gets d + 1$
        \EndWhile
        
        \Comment{Check for successful compilation and length reduction}
        \If{$success$ \textbf{and} $\textsc{Length}(P_{cand}) < \textsc{Length}(P_{curr})$}
            \State $P_{curr} \gets P_{cand}$
            \State $History \gets History \cup \{(step, \text{Success})\}$
            \State $step\_applied \gets \textbf{True}$
            \State \textbf{break} \Comment{Break inner loop to dynamically replan with updated proof}
        \EndIf
    \EndFor
    
    \If{\textbf{not} $step\_applied$}
        \State $History \gets History \cup \{(Plan, \text{Failed})\}$
    \EndIf
    \State $steps \gets steps + 1$
\EndWhile
\State \Return $P_{curr}$
\end{algorithmic}
\end{algorithm}

\newpage

\section{Prompts}
\label{sec:prompts}

\subsection{Planner Prompt}

\begin{promptbox}
Your task is to analyze a given Lean 4 proof and create a structured plan for optimizing and shortening it, while maintaining the correctness.

## Instructions

Analyze the proof and identify regions (or entire proof) that can be optimized and simplified. For each optimization opportunity:

1. **Identify the line range**: ...
2. **Provide a title**: ...
3. **Provide potential reduction**: ...
4. **Determine an optimization strategy**: ...

## Inputs

1. A correct Lean 4 statement and proof source code.
2. Elaborated signature.
3. Doc string from the source if it exists.
4. All the dependencies used in the statement and proof...
5. A list of retrieved optimization strategies from a strategy index...

## Retrieved Optimization Strategies
...

## Output Format

Output the plans using the following JSON format, wrapped in a single ```json ``` tags. Ensure the output is valid JSON.

[
  {
    "line_start": X,
    "line_end": Y,
    "title": "the strategy name",
    "reduction": "high, medium, or low",
    "description": "Detailed description of the optimization strategy ..."
  }
]

Rules about the ordering of the output plans:
1) Sort plans primarily from top to bottom of the proof (increasing line numbers).
2) Overlaps of optimization regions are allowed. If two plans overlap in location, put the plan with the MOST significant potential reduction first.
3) If two plans have the same overlap and same potential impact, keep top-to-bottom ordering.

Examples:
...

Here is the proof to optimize:
...

Theorem's elaborated signature:
...

Theorem's documentation:
...

Information about dependencies used in the proof:
...

Now analyze the proof and generate your optimization plans.
\end{promptbox}

\subsection{Refactor Prompt}
\begin{promptbox}
You are an expert Lean 4 proof optimizer. Your task is to execute a specific optimization plan on a proof, modifying the targeted section while ensuring the proof remains correct.

## Current Proof
...

## Theorem's Elaborated Signature
...

## Documentation
...

## Dependencies
...

## Optimization Plan

**Target Lines:**: ...
**Title:**: ...
**Description:**: ...

## Instructions

1. **Focus on the targeted section**: ...
2. **Preserve correctness**: ...
3. **Do NOT modify the theorem statement**: ...
4. **Aim for shorter proof**: ...
5. **Output the complete theorem and proof**: ...

Before providing the code, briefly explain what changes you're making and why.

Then output the complete optimized Lean 4 code wrapped in tags:

```lean4
<your optimized complete theorem and proof here>
```
\end{promptbox}

\subsection{Debugger Prompt}
\begin{promptbox}
The proof (Round {{ prev_round_num }}) is not correct. Following is the compilation error message, where we use <error></error> to signal the position of the error.

Error message:
...

Before producing the Lean 4 code that fixes the error, provide a detailed analysis of the error message. Focus on fixing the compilation errors, and try your best to preserve the proof optimization that was previously done. Do not revert to the original unoptimized proof.

You must wrap the entire Lean 4 theorem and proof in tags like:

```lean4
<your corrected proof here>
```
\end{promptbox}

\section{Test Time Scaling Curves}
\label{sec:tts_figures}

\begin{figure}[H] % Standard single-column figure
    \centering
    % Top Left
    \begin{subfigure}[b]{0.48\linewidth}
        \centering
        \includegraphics[width=\textwidth]{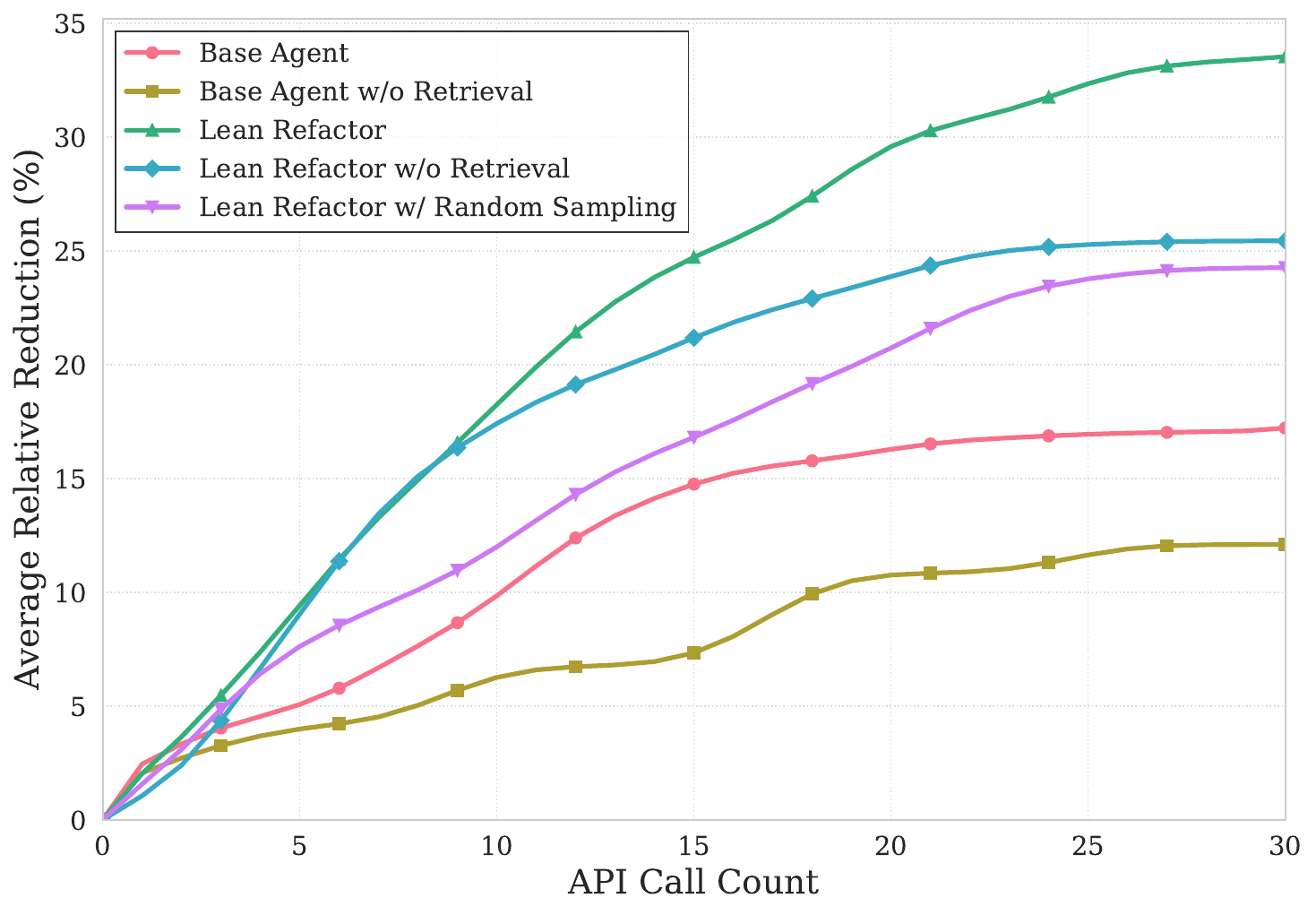}
        \caption{Improvements on Analysis Dataset}
        \label{fig:fig2_a}
    \end{subfigure}
    \hfill
    % Top Right
    \begin{subfigure}[b]{0.48\linewidth}
        \centering
        \includegraphics[width=\textwidth]{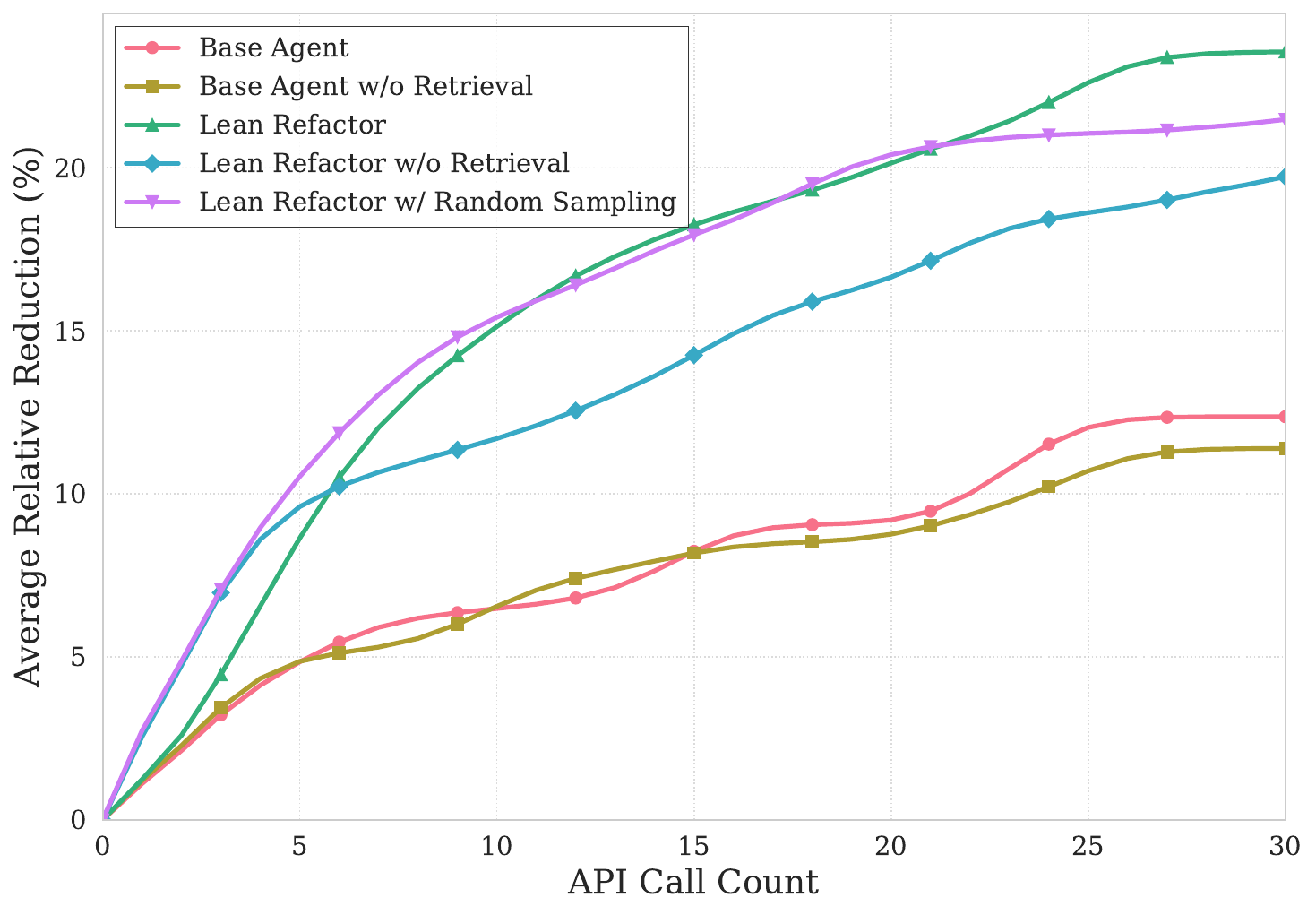}
        \caption{Improvements on FLT Dataset}
        \label{fig:fig2_b}
    \end{subfigure}
    
    \vspace{1em} % Adds vertical space between the top and bottom rows
    
    % Bottom Left
    \begin{subfigure}[b]{0.48\linewidth}
        \centering
        \includegraphics[width=\textwidth]{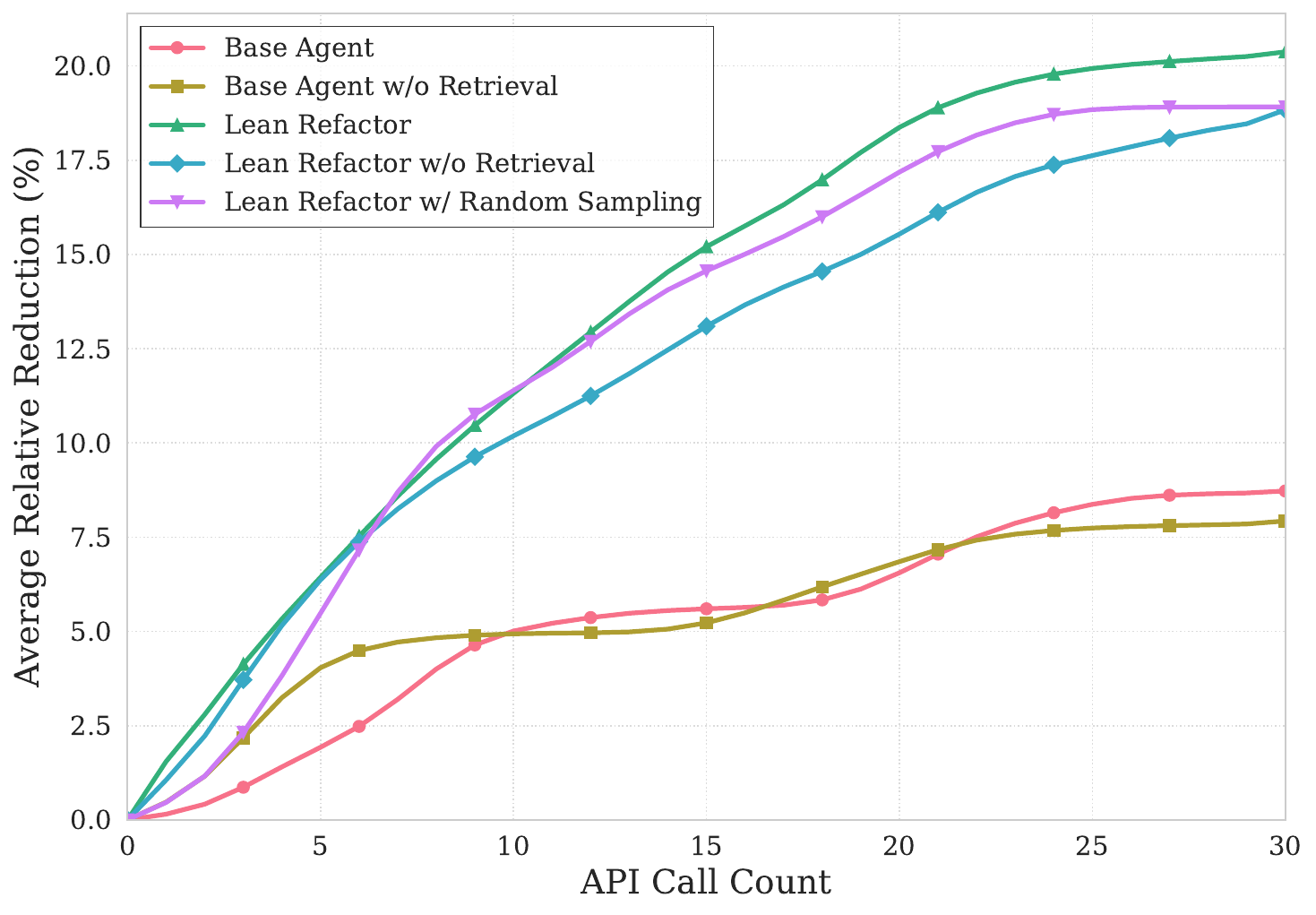}
        \caption{Improvements on PFR Datasets}
        \label{fig:fig2_c}
    \end{subfigure}
    \hfill
    % Bottom Right
    \begin{subfigure}[b]{0.48\linewidth}
        \centering
        \includegraphics[width=\textwidth]{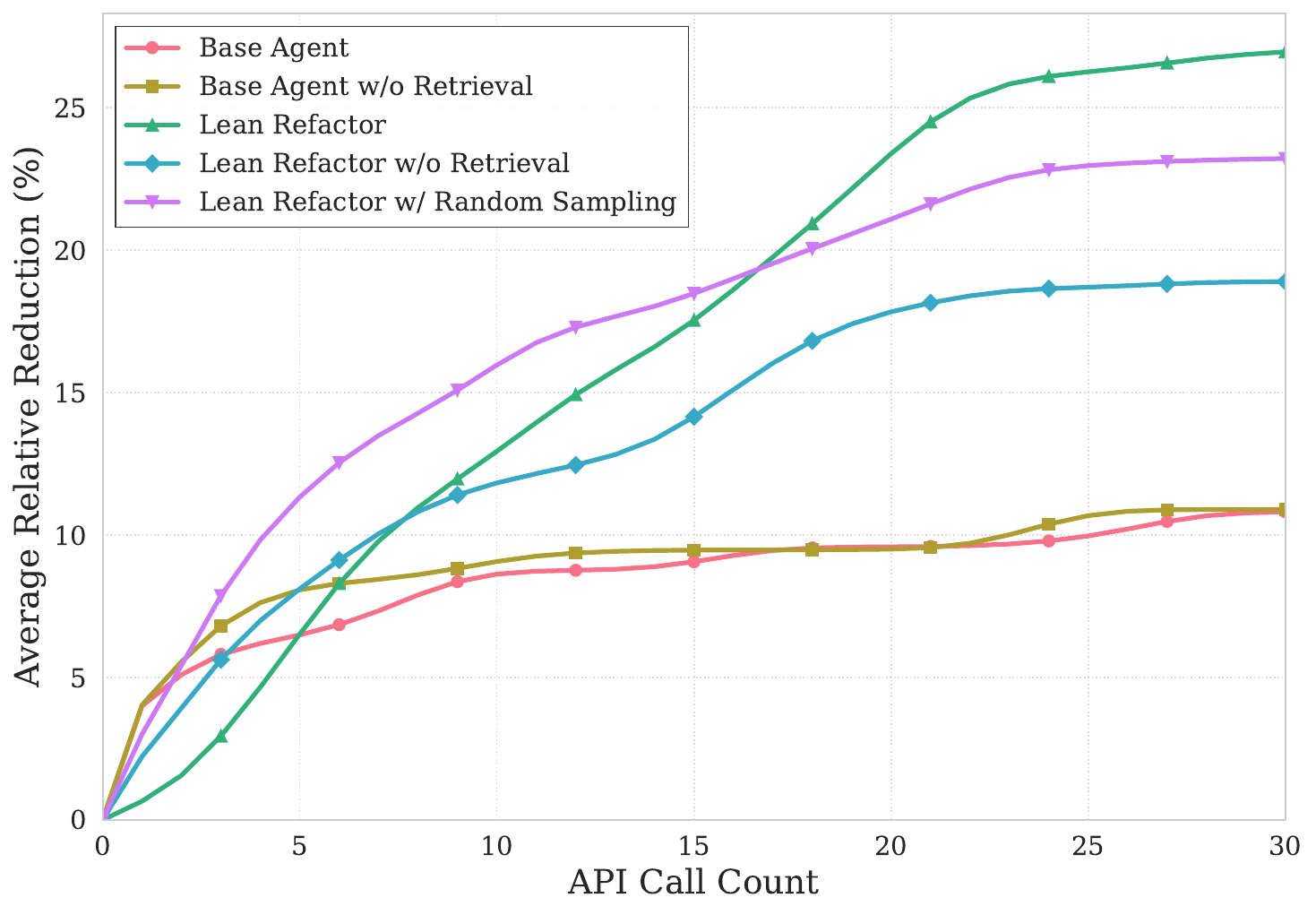}
        \caption{PhysLean Improvements}
        \label{fig:fig2_d}
    \end{subfigure}
    
    \caption{\textbf{Average relative length reduction across research-level datasets over successive API calls.} We compare the base agent against our proposed Lean Refactor agent, alongside ablation variants (without retrieval and with random sampling). The Lean Refactor agent converges to a higher relative reduction compared to all other variants.}
    \label{fig:four_images}
\end{figure}

\begin{figure}[H] % The asterisk (*) makes it span both columns
    \centering
    % First Subfigure
    \begin{subfigure}[b]{0.32\textwidth}
        \centering
        \includegraphics[width=\textwidth]{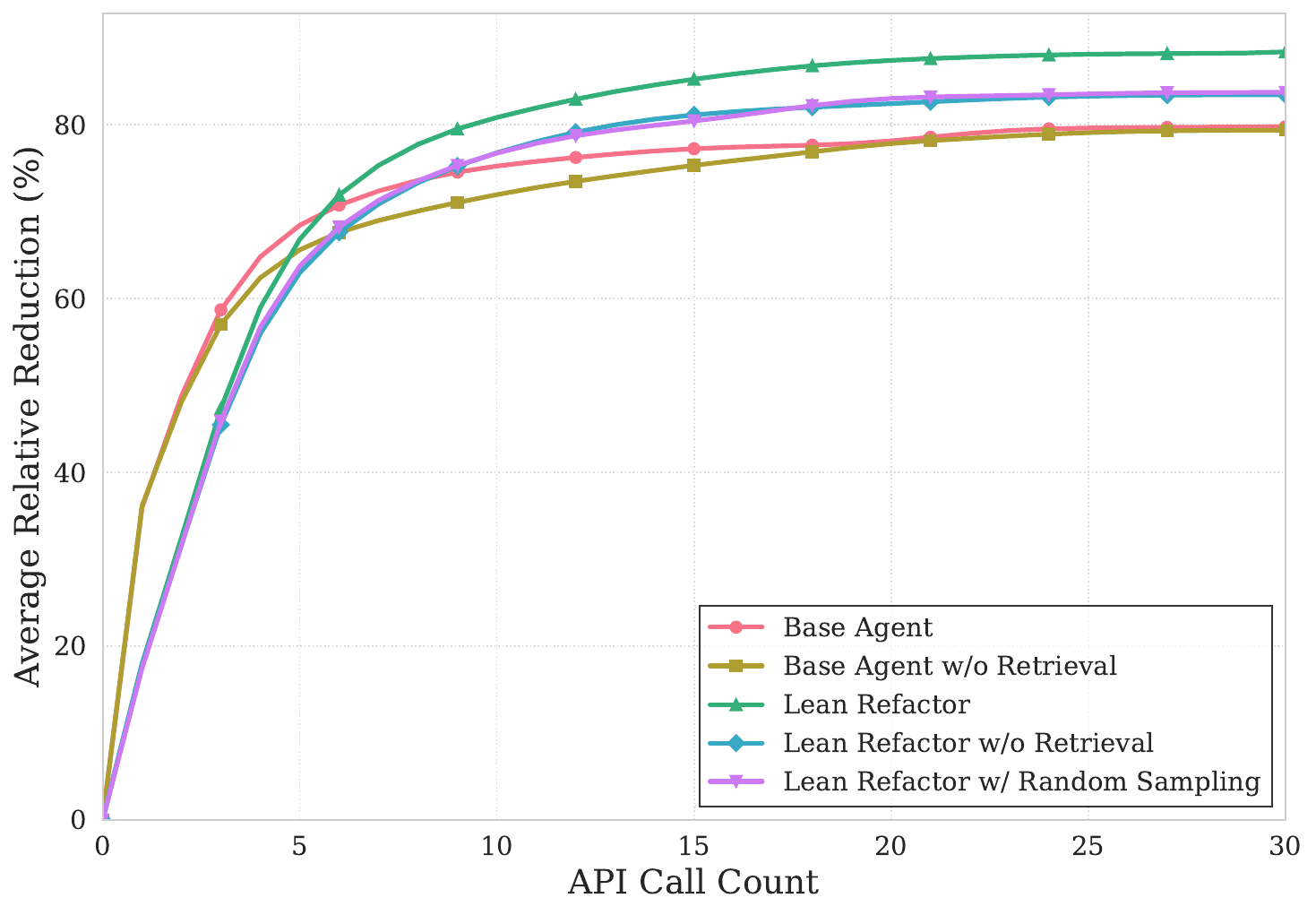} % Replace with your file
        \caption{miniF2F Improvements}
        \label{fig:fig1_a}
    \end{subfigure}
    \hfill % Adds flexible spacing between figures
    % Second Subfigure
    \begin{subfigure}[b]{0.32\textwidth}
        \centering
        \includegraphics[width=\textwidth]{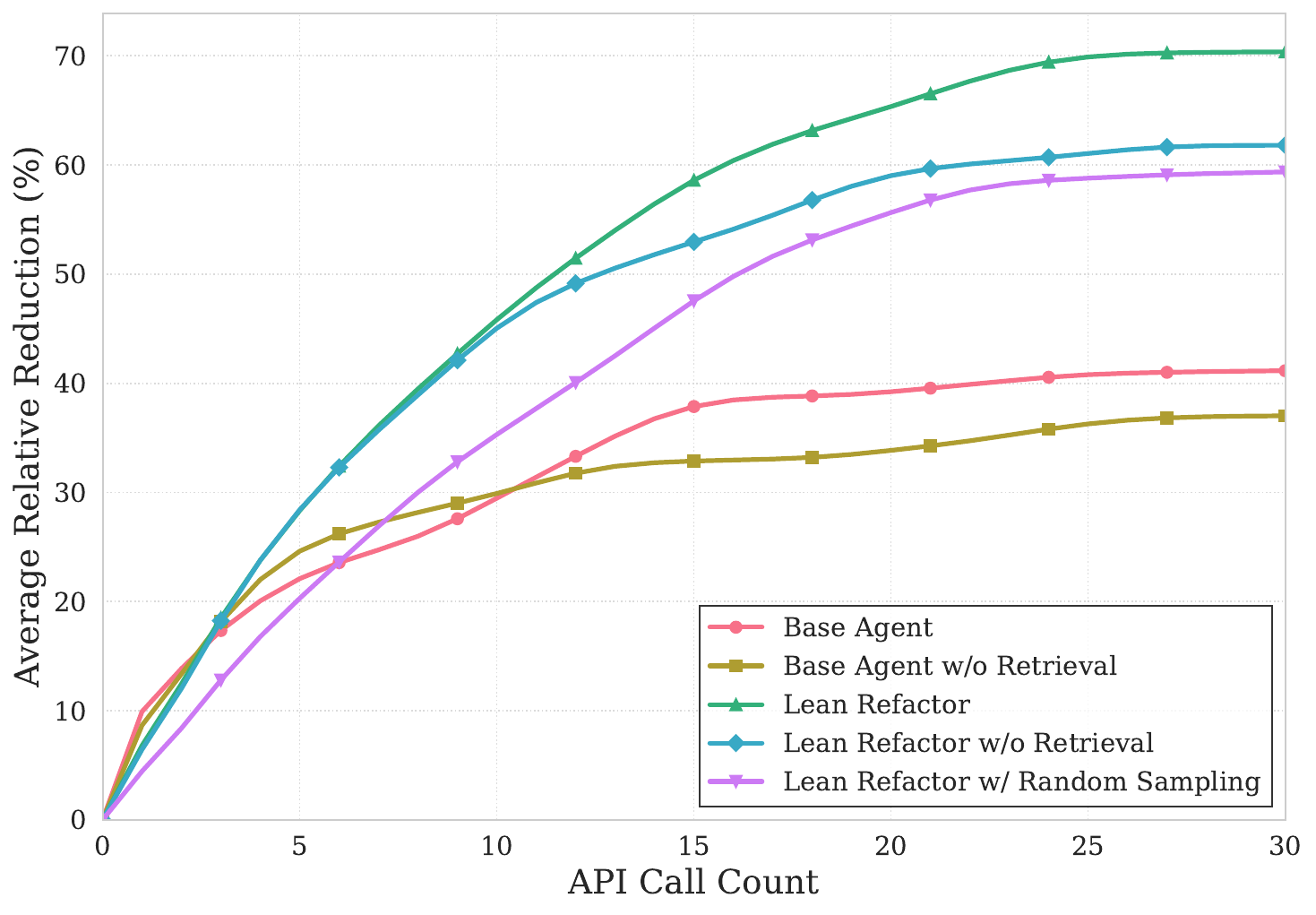}
        \caption{PutnamBench Improvements}
        \label{fig:fig1_b}
    \end{subfigure}
    \hfill
    % Third Subfigure
    \begin{subfigure}[b]{0.32\textwidth}
        \centering
        \includegraphics[width=\textwidth]{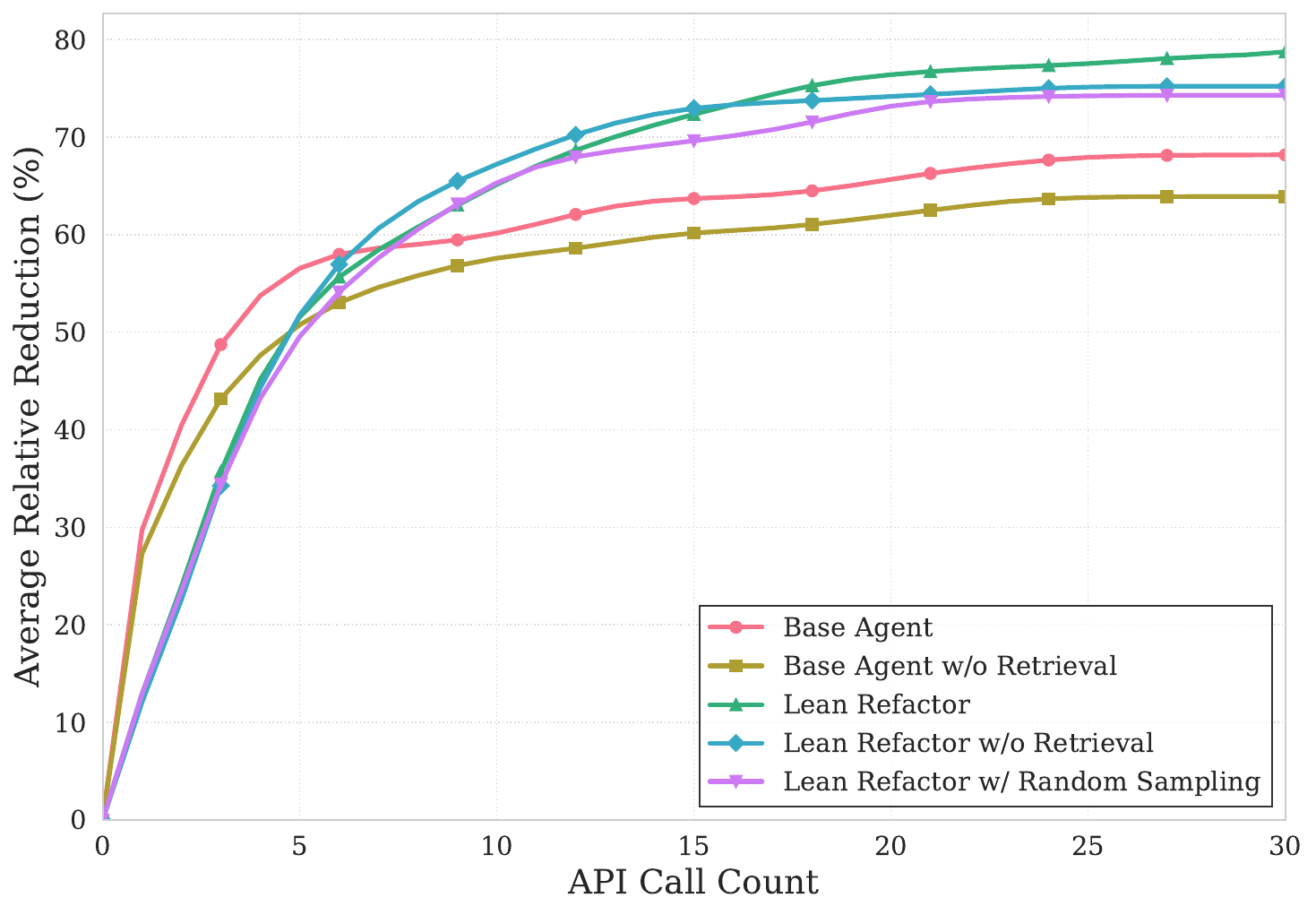}
        \caption{Putnam 2025 Improvements}
        \label{fig:fig1_c}
    \end{subfigure}
    
    \caption{\textbf{Average relative length reduction across competition-style datasets over successive API calls.} We compare the base agent against our proposed Lean Refactor agent, alongside ablation variants (without retrieval and with random sampling). The Lean Refactor agent converges to a higher relative reduction compared to all other variants.}
    \label{fig:competition_improvements}
\end{figure}

% \newpage
% \input{checklist.tex} for neurips !!!

\end{document}